\documentclass[twocolumn,twoside]{IEEEtran}

\ifCLASSINFOpdf

\else

\fi

\ifCLASSOPTIONcompsoc
    \usepackage[caption=false, font=normalsize, labelfont=sf, textfont=sf]{subfig}
\else
    \usepackage[caption=false, font=normalsize]{subfig}
\fi
\usepackage{lipsum}%
\usepackage[dvipsnames]{xcolor}
\usepackage{algorithm,algorithmic}
\usepackage{balance}
\usepackage{multicol}   
\usepackage{cite}
\usepackage{gensymb}
\usepackage{multirow}
\usepackage{graphics}
\usepackage{epsfig}
\usepackage{graphicx}
\usepackage{epstopdf}
\usepackage{textcomp}
\usepackage{amsmath}
\usepackage{mathtools}
\usepackage{filecontents}
\usepackage{lipsum,color}
\usepackage{amssymb}
\usepackage{float}
\usepackage{colortbl} 
\usepackage{times} 
\usepackage{amsthm}  

\usepackage{array}
\usepackage{tabularx}
\usepackage{makecell}

\usepackage{amsfonts}
\usepackage{hyperref}

\definecolor{royalblue}{RGB}{65,55,255}
\newcommand{\rev}[1]{{\color{black}#1}}
\usepackage{booktabs}  

\theoremstyle{break}

\begin{document}

\title{Analytical Channel Modeling and Stability Aware Optimization of Optical Inter Satellite Links}

\author{Hossein~Safi, {\it Member,~IEEE}, Ziheng Wang, Stijn Mast, \\ Harald~Haas, {\it Fellow,~IEEE}, and Iman~Tavakkolnia, {\it Senior Member,~IEEE}
	
\thanks{H. Safi, Z. Wang, H. Haas and I. Tavakkolnia are with LiFi Research and Development Centre, Department of Engineering, Cambridge University, Cambridge CB3 0FA, UK. (e-mails: {hs905, zw475, huh21, it360}@cam.ac.uk). S. Mast is with European Space Agency (email:stijn.mast@esa.int). }}

\maketitle
\begin{abstract}
Optical inter‑satellite links (OISLs) are key enablers for high‑capacity space networks and next‑generation satellite constellations. However, their extreme directionality makes link reliability highly sensitive to platform‑induced pointing jitter, which causes random misalignment between the transmitter and receiver beams. In this paper, we develop a tractable closed‑form statistical channel model for point‑to‑point OISLs subject to independent pointing errors at both terminals. Accurate Gaussian main‑lobe approximations are applied to the transmitter far‑field pattern and receiver coupling efficiency. This transforms the diffraction‑based channel response into closed‑form expressions for the channel‑gain distribution, outage probability, and ergodic capacity. The analytical results are validated through Monte Carlo simulations and used to study the impact of terminal stability, beam divergence, and link margin on OISL performance. The results show that outage probability is governed by the weaker terminal in terms of pointing stability, while improving only the stronger terminal provides minimal additional benefit. In contrast, the ergodic‑capacity penalty depends on the combined stability of both terminals, revealing a fundamental distinction between reliability and throughput metrics. The proposed framework provides practical design guidelines for selecting beam parameters and specifying pointing and tracking requirements under varying levels of platform instability.
\end{abstract}
\begin{IEEEkeywords}
Channel modeling,  free-space optics, non-terrestrial networks, OISL, pointing errors.
\end{IEEEkeywords}
\IEEEpeerreviewmaketitle

\section{Introduction}

\subsection{Background and Motivation}

The rapid deployment of large-scale low Earth orbit (LEO) satellite constellations is reshaping the architecture of global communication networks and accelerating the transition from conventional radio-frequency (RF) backbones toward optical inter-satellite links (OISLs) \cite{non-terrestrial1, non-terrestrial2, non-terrestrial3, OISL-1, optical_ntn}. By exploiting the vast unlicensed optical spectrum, OISLs enable multi-gigabit to terabit-per-second connectivity with significantly reduced size, weight, and power requirements compared to RF inter-satellite links. The inherently narrow beam divergence of laser transmission further provides enhanced security and low inter-link interference, which is critical for densely populated constellations \cite{OISL-1}. Beyond capacity gains, OISLs offer the potential for ultra-low latency global connectivity, as optical signals propagate faster in vacuum than in terrestrial fiber \cite{optical_channel}, allowing satellite networks to outperform ground-based long-haul infrastructure for intercontinental data transport \cite{OISL_routing, inter_satellite_links}. These advantages have driven the transition of OISLs from early experimental demonstrations to operational deployment in emerging commercial constellations.

Recent advances in optical terminal design, coherent modulation formats, and digital signal processing have further expanded the achievable data rates and link distances of OISLs \cite{advances-1, advances-2}. However, the extreme directionality required to overcome free-space path loss over inter-satellite distances renders these links highly sensitive to mechanical vibrations, attitude control errors, and residual tracking inaccuracies \cite{advances-3}. 
In practice, pointing errors arise not only from high-frequency platform jitter but also from slowly varying bias terms caused by thermo-mechanical drift and calibration imperfections in the pointing, acquisition, and tracking (PAT) subsystem. These bias errors are typically estimated and compensated through periodic calibration and closed-loop tracking, leaving a residual stochastic misalignment that fluctuates around the corrected boresight direction. As atmospheric effects are negligible in space, these residual platform-induced pointing errors become the dominant physical-layer impairment governing link reliability and throughput. Consequently, OISL performance is fundamentally constrained by the stability of the PAT subsystems and by the interaction between optical beam divergence, receiver field-of-view (FOV), and platform jitter \cite{PAT_effect1, PAT_effect2}.

The statistical impact of pointing errors has been widely studied in terrestrial and non-terrestrial free-space optical systems \cite{PAT_effect1, channel_model, pointing-1, pointing-2, channel_model2, channel_model3}. Prior analytical models typically adopt Gaussian beam profiles and Rayleigh-distributed radial pointing errors, enabling closed-form performance expressions under simplified assumptions. Existing treatments of inter-satellite laser links have further derived tight bounds or closed-form expressions for error performance under misalignment at a single terminal and specific detection schemes \cite{channel_model4}. However, most of the available literature assumes perfect tracking at the opposite end or treats one terminal as deterministic, an assumption that is reasonable for asymmetric scenarios such as ground-to-air or deep-space downlinks but becomes increasingly unrealistic for symmetric OISLs in which both satellites employ similar optical terminals and operate under comparable disturbance environments. Although the physically correct end-to-end channel model for OISLs must account for the composite effect of two independent misalignment processes at the transmitter and receiver, whose impacts combine multiplicatively on the received power \cite{channel_model5,pointing-3,OISL-retro}, existing works do not provide a closed-form statistical characterization of the resulting channel gain distribution under joint transmitter and receiver pointing errors. Instead, most treatments rely on numerical integration or Monte Carlo simulation. Instead, they largely rely on numerical integration or Monte Carlo simulation, which limits analytical insight and complicates system-level optimization. In particular, existing models do not provide tractable insight into how imbalances between terminal stabilities shape outage behavior or into how beamwidth and receiver FOV should be optimized for link dimensioning.

Moreover, current design methodologies for OISLs remain largely power-centric, focusing on increasing transmit power or antenna gain to combat link outages \cite{PAT_effect1, channel_model5}. This approach neglects the intrinsic coupling between beam divergence, pointing robustness, and physical payload constraints, and fails to capture the diminishing returns of unilateral stability enhancement at one terminal. As demonstrated in related aerial and optical communication systems \cite{pointing-2, channel_model2, channel_model3}, link reliability in the presence of platform fluctuations is governed by a stability-centric tradeoff between beamwidth, pointing jitter, and geometric gain, rather than by transmit power alone.

Motivated by these gaps, this paper develops a unified analytical framework for OISLs subject to independent pointing errors at both terminals. By transforming the diffraction-based channel response into tractable statistical expressions, the proposed model enables closed-form evaluation of outage probability and ergodic capacity across both stable and jitter-limited regimes. This framework reveals fundamental stability-driven design principles for OISLs, quantifies the performance limits imposed by the weaker terminal, and provides explicit guidelines for joint optimization of beam divergence, receiver FOV, and link margin under realistic payload and PAT constraints. The resulting analytical tools offer a direct means to translate pointing stability specifications into availability and throughput targets for next-generation optical satellite networks.

\subsection{Major Contributions and Novelty}

The main contribution of this paper is a tractable, physically grounded statistical framework for characterizing OISL reliability and throughput under platform‑induced pointing jitter. Particularly, by combining accurate Gaussian main-lobe approximations of the transmitter far-field intensity pattern and the receiver coupling efficiency with a Rayleigh jitter model, we derive closed-form expressions that enable direct performance evaluation and system-level optimization without resorting to time-consuming simulations. 

In summary, the specific contributions of this work are listed as follows:
\begin{itemize}
	\item We derive exact closed-form expressions for the probability density function (PDF) of the instantaneous end-to-end channel gain under independent Rayleigh-distributed pointing errors, parametrized by two dimensionless stability parameters $\phi_{\text{tx}}$ and $\phi_{\text{rx}}$ that capture the robustness of the transmitter beamwidth and receiver FOV to pointing jitter.
	\item Based on the derived channel model, we obtain closed-form expressions for the outage probability for arbitrary stability values and \rev{derive a high-SNR asymptotic ergodic-capacity penalty} that quantifies the throughput degradation due to pointing jitter for both coherent and direct-detection OISL. \rev{The corresponding exact finite-SNR ergodic capacity is also expressed as a single one-dimensional integral over the closed-form gain PDF}.
	\item Through high-margin asymptotic analysis, we establish a weakest-link principle, showing that the outage decay is governed by the less stable terminal, while improvements at the stronger terminal yield only limited gains through a power offset. We further show that outage and ergodic capacity depend on fundamentally different stability metrics.
	\item We compare the Gaussian beam and coupling approximations with exact diffraction-based and Airy-pattern-based calculations. The results confirm high accuracy within the nominal pointing accuracy range of operational OISLs.
	\item We translate the analytical results into explicit design guidelines and propose an iterative link design procedure that jointly optimizes beam divergence, receiver FOV, and link margin under physical aperture constraints and the weakest-link principle.
	\item We further extend the analysis to bi-directional OISLs and heterogeneous terminals, deriving system-level outage and symmetric-rate limits, and providing explicit beamwidth and power-scaling rules for stability balancing under asymmetric pointing jitter.
\end{itemize}
\rev{A dedicated comparative table positioning the present framework against prior pointing-error and channel-modeling works is provided in Table~\ref{tab:novelty}.}

\begin{table*}[!t]
\rev{
\centering
\caption{Positioning of the present work against representative prior OISL and FSO pointing-error/channel-modeling literature.}
\label{tab:novelty}
\scriptsize
\renewcommand{\arraystretch}{1.18}
\setlength{\tabcolsep}{2.5pt}
\begin{tabularx}{\textwidth}{
p{2.25cm}
p{1.45cm}
p{1.55cm}
p{1.45cm}
p{1.55cm}
p{1.35cm}
X
p{1.35cm}
p{1.65cm}}
\toprule
\textbf{Reference} &
\makecell[l]{\textbf{Pointing}\\\textbf{model}} &
\makecell[l]{\textbf{Beam}\\\textbf{model}} &
\makecell[l]{\textbf{Main}\\\textbf{metric}} &
\makecell[l]{\textbf{Analysis}\\\textbf{type}} &
\makecell[l]{\textbf{Link}\\\textbf{type}} &
\textbf{Optimized or studied variables} &
\makecell[l]{\textbf{Acq.}\\\textbf{modeled}} &
\makecell[l]{\textbf{Finite-SNR}\\\textbf{capacity}} \\
\midrule

\cite{pointing-1} &
TX wander + finite RX aperture &
Gaussian &
Outage capacity &
Closed form &
Terrestrial FSO &
Beam width and code rate &
No &
Binary-input AWGN outage \\

\cite{pointing-2} &
Mostly TX; both discussed &
Various &
BER, outage, capacity &
Survey &
Mostly uni. &
Reviewed, not optimized &
Reviewed &
No new derivation \\

\cite{channel_model2} &
TX only for hovering UAV &
Gaussian &
BER, outage &
Closed form &
UAV-FSO &
Divergence and RX aperture &
No &
No \\

\cite{channel_model3} &
TX only &
Gaussian &
BER, outage &
Closed form &
Ground-HAP &
Divergence and transmit power &
No &
No \\

\cite{channel_model4} &
Composite pointing term &
Gaussian, large-detector limit &
Ergodic rate &
Closed form &
Relay-assisted uni. &
Laser frequency and relay number &
No &
Large-detector limit \\

\cite{channel_model5} &
TX only; TX+LO for Het. &
Gaussian; Airy also discussed &
BER &
Closed form &
Uni. &
TX/RX antenna gain and laser power &
No &
No \\

\cite{pointing-3} &
TX only; wide RX FOV &
Gaussian &
BER &
Closed form &
Uni. &
TX antenna gain &
No &
No \\

\cite{OISL-retro} &
TX + retro + RX &
Airy + Gaussian &
BER, outage, AIR &
Monte Carlo &
Round-trip retroreflector &
Divergence, aperture, FOV, $N_{\mathrm{EXT}}$ &
Implicit &
AIR via MC \\

\cite{jsac-channel} &
TX pointing + beam wander &
Truncated Gaussian &
Outage &
Closed form &
Uplink/downlink &
Telescope diameters and AO compensation &
Tracking discussed &
No \\

\cite{pointing-rayleigh} &
TX only &
Gaussian &
Acquisition probability &
Closed form &
ISL &
Beacon divergence and scan strategy &
Yes &
No \\

\cite{pointing-rayleigh2} &
TX only &
Gaussian &
BER &
Closed form &
Uni. &
Divergence-to-jitter ratio &
No &
No \\

\midrule
\textbf{This work} &
\textbf{TX + RX, two independent Rayleigh terms} &
\textbf{Gaussian; validated against Airy/Klein--Degnan} &
\textbf{Outage, ergodic-capacity penalty} &
\textbf{Closed-form PDF/CDF + 1-D capacity integral} &
\textbf{Uni. and bi.} &
\textbf{Power, divergence, and FOV through constrained optimization} &
\textbf{No} &
\textbf{Exact finite-SNR integral} \\

\bottomrule
\end{tabularx}
\vspace{-1mm}
\begin{flushleft}
\footnotesize{
TX: transmitter; RX: receiver; Het.: heterodyne; FOV: field of view; AO: adaptive optics; AIR: achievable information rate; MC: Monte Carlo; Uni.: unidirectional; Acq.: acquisition.}
\end{flushleft}
}
\end{table*}

The remainder of this paper is organized as follows. Section~\ref{sec:sysmodel} presents the system model and formulates the deterministic and stochastic components of the channel gain. Section~\ref{III-stat} derives the statistical characterization of the channel and validates the underlying Gaussian approximations. Section~\ref{sec:outage} presents the closed-form outage probability analysis, and Section~\ref{sec:asymptotic} examines the high-margin asymptotic regime to extract the decay exponent and power offset, and extends the framework to bi-directional links and asymmetric terminals. Section~\ref{sec:numerical} provides numerical results and performance evaluation. Finally, Section~\ref{sec:conclusion} concludes the paper and discusses future research directions.

\section{System Model and Channel Gain Analysis}
\label{sec:sysmodel}

The performance of an OISL is fundamentally governed by the stability of highly directional optical beams over long inter satellite distances.  This section develops a tractable end to end channel model for a point to point OISL between two satellite platforms separated by a distance $z$ and operating at wavelength $\lambda$, which forms the basis for the subsequent statistical and performance analysis.

\subsection{Transmitter Model: Truncated Gaussian Beam and Far-Field Pattern}
We assume that the transmitter utilizes a telescope with effective aperture of diameter $D_\text{tx}$ to project a collimated laser beam. In practical OISL systems, the beam profile is characterized as a truncated Gaussian due to the finite size of the transmit telescope \cite{OISL-retro,jsac-channel}. We define the truncation ratio as $\alpha_0 = a_\text{tx}/w_0$, where $a_\text{tx} = D_\text{tx}/2$ is the aperture radius and $w_0$ is the Gaussian beam waist at the transmitter. The parameter $\alpha_0$ is typically chosen to maximize the on-axis intensity in the far-field, balancing diffraction effects against aperture truncation \cite{jsac-channel}.

The instantaneous transmitter gain in the direction of the receiver, accounting for pointing error $\theta_\text{tx}$ (the off-boresight angle), is given by \cite[(6)]{Antena-gain}
\begin{equation}
	G_\text{tx}(\theta_\text{tx}) = G_\text{tx0} \cdot L_\text{tx}(\theta_\text{tx}),
\end{equation}
where $G_\text{tx0} = 4\pi A_\text{tx} / \lambda^2$ is the peak on-axis gain with $A_\text{tx} = \pi a_\text{tx}^2$, and $L_\text{tx}(\theta_\text{tx})$ is the normalized far-field intensity pattern (FFIP). The FFIP, also known as the transmitter efficiency factor, is obtained from the Fraunhofer diffraction integral of the truncated Gaussian source as \cite[(7)]{Antena-gain}
\begin{equation}
	\label{eq:ffip}
	L_\text{tx}(\theta_\text{tx}) = {2\alpha_0^2 \left| \int_{\gamma_o^2}^1 \exp(-\alpha_0^2 u) J_0(X\sqrt{u}) \, du \right|^2}.
\end{equation}
Here, $X = (\pi D_\text{tx}/\lambda) \sin(\theta_\text{tx}) \approx (\pi D_\text{tx}/\lambda) \theta_\text{tx}$ for small angles, and $J_0(\cdot)$ is the zero-order Bessel function of the first kind. Furthermore, in \eqref{eq:ffip}, \rev{$\gamma_o = a_\text{obs}/a_\text{tx}\in[0,1)$ denotes the {obscuration ratio}, defined as the radius of the central obscuration relative to the transmit aperture radius \cite{Antena-gain}. The two dimensionless parameters governing the transmitter efficiency factor are therefore the truncation ratio $\alpha_0 = a_\text{tx}/w_0$ and the obscuration ratio $\gamma_o$.} It is shown that \rev{the on-axis efficiency $L_\text{tx}(0)$} is maximized when $\alpha_{0} \approx 1.12 - 1.30\,\gamma_o^{2} + 2.12\,\gamma_o^{4}$ \cite{Antena-gain,jsac-channel}\rev{, where the right-hand side is a function of the obscuration ratio $\gamma_o$ evaluated under the optimum-truncation condition}.

\rev{The function $L_\text{tx}(\theta_\text{tx})$ defined in \eqref{eq:ffip} is the Klein--Degnan transmitter efficiency factor \cite{Antena-gain}, which incorporates the deterministic truncation and obscuration losses. It is therefore {not} normalized to unity at boresight. In particular, $L_\text{tx}(0)\le 1$, with strict inequality whenever $\alpha_0$ is finite or $\gamma_o>0$. For the unobscured optimally-truncated case ($\gamma_o=0$, $\alpha_0=1.12$), $L_\text{tx}(0)=2(1-e^{-\alpha_0^2})^2/\alpha_0^2 \approx 0.815$, i.e., approximately $-0.89$~dB of combined truncation/obscuration loss relative to a uniformly illuminated unobscured aperture of the same area. To separate the deterministic on-axis efficiency from the pointing-induced stochastic loss, we define the transmitter taper efficiency
\begin{equation}
	\eta_T \triangleq L_\text{tx}(0), \qquad
	\widetilde{L}_\text{tx}(\theta_\text{tx}) \triangleq \frac{L_\text{tx}(\theta_\text{tx})}{L_\text{tx}(0)},
\end{equation}
so that $\widetilde{L}_\text{tx}(0)=1$ by construction. The actual peak on-axis transmitter gain is then $G_\text{tx0}\,\eta_T$, with $G_\text{tx0}=4\pi A_\text{tx}/\lambda^2$ retained as a uniform-aperture reference gain. The deterministic factor $\eta_T$ is absorbed into the link budget through the peak channel gain $G_\text{peak}$ defined in Section~\ref{sec:sysmodel}-D below.}

\subsection{Free-Space Path Loss}
The isotropic spreading of the wavefront over the vacuum channel is characterized by the free-space path loss (FSPL)
\begin{equation}
	L_s(z) = \left( \frac{\lambda}{4\pi z} \right)^2.
\end{equation}
For long-range intersatellite distances (e.g., hundreds to thousands of kilometers), this geometric attenuation is substantial, necessitating high-gain optical assemblies to ensure sufficient received power for link closure.

\subsection{Receiver Model}
%
%
%
\subsubsection{Coupling Efficiency}
The receiver telescope, with aperture diameter $D_\text{rx}$, collects the incident light and focuses it onto a photodetector of finite size. The receiver's angular acceptance is defined by its FOV, $\theta_{\text{fov}}$, which determines the region over which the detector can successfully capture the focused spot. Similar to the transmitter side, the receiver gain is expressed as 
\begin{equation}
	G_\text{rx}(\theta_\text{rx}) = G_\text{rx0} \cdot L_\text{rx}(\theta_\text{rx}), 
\end{equation}
where $G_\text{rx0} = 4\pi A_\text{rx} / \lambda^2$ and $\theta_\text{rx}$ models angular misalignment of the receiver optical axis relative to the incident beam direction, i.e., the residual jitter after acquisition and tracking. 

Particularly, we assume nominal boresight alignment between the transmitter and receiver, corresponding to the steady-state tracking regime after successful acquisition. In this regime, the pointing control loop maintains alignment with only small residual errors. The orthogonal angular jitter components (e.g., azimuth and elevation) are modeled as zero-mean random processes, while the radial pointing error magnitude $\theta_\text{rx}$ represents the resulting misalignment amplitude.
Systematic boresight offsets and higher order aberrations are not considered and are absorbed into the deterministic efficiency factors.
Accordingly, the coupling efficiency $L_\text{rx}(\theta_\text{rx})$ represents the fraction of the focused power (modeled by the Airy diffraction pattern for a circular aperture) that falls within the detector area. For a misalignment error $\theta_\text{rx}$, this is given by \cite{OISL-retro}
\begin{equation}
	\label{eq:coupling}
	L_\text{rx}(\theta_\text{rx}) = \frac{\iint_{\mathcal{A}_{\text{det}}} \mathcal{P}_{\text{Airy}}(\vec{r} - \vec{d}_\text{rx}) \, d\vec{r}}{\iint_{\infty} \mathcal{P}_{\text{Airy}}(\vec{r}) \, d\vec{r}},
\end{equation}
where $\vec{d}_\text{rx} \approx f \cdot \vec{\theta}_\text{rx}$ is the displacement in the focal plane due to the pointing error, $f$ is the receiver focal length, and $\mathcal{P}_{\text{Airy}}$ is the Airy intensity pattern. \rev{Analogously to the transmitter side, $L_\text{rx}(\theta_\text{rx})$ is an absolute coupling efficiency that includes the on-axis Airy-to-detector spillover. Its zero-misalignment value $L_\text{rx}(0)$ is in general strictly less than unity and depends on the ratio of the detector radius to the Airy radius. We therefore define the receiver on-axis coupling efficiency $\eta_R\triangleq L_\text{rx}(0)$ and the normalized stochastic receiver loss $\widetilde{L}_\text{rx}(\theta_\text{rx})\triangleq L_\text{rx}(\theta_\text{rx})/L_\text{rx}(0)$, with $\widetilde{L}_\text{rx}(0)=1$ by construction. The deterministic factor $\eta_R$ is absorbed into the peak channel gain $G_\text{peak}$ (defined in Section~\ref{sec:sysmodel}-D), while $\widetilde{L}_\text{rx}$ is the stochastic loss factor that enters the closed-form statistical analysis of Section~\ref{III-stat}.}

The focal-plane irradiance of a diffraction-limited circular receiver telescope is given by the Airy intensity pattern. The coupling efficiency into a finite-area photodetector is therefore equal to the fraction of the Airy pattern captured by the detector active area after a focal-plane displacement $\vec{d}_\text{rx}$ caused by angular misalignment $\vec{\theta}_\text{rx}$. Although the exact coupling efficiency can be computed numerically from the displaced Airy pattern, this expression does not admit a closed-form solution and complicates statistical performance analysis.

In the steady-state tracking regime relevant to practical OISLs, this error is typically small compared with the angular acceptance of the receiver \cite{jsac-channel,rx-fov-approx}, and the dominant contribution to coupling loss arises from the roll-off of the main lobe of the Airy pattern. In this regime, the central lobe of the Airy pattern is well approximated by a two-dimensional Gaussian. Consequently, the coupling efficiency into a photodetector of finite radius can be reasonably modeled as an exponential function of the squared pointing error. This approximation, which will be expressed mathematically in the sequel, is accurate within the small-misalignment regime characteristic of closed-loop pointing and tracking in OISL systems when the link is established \cite{deep-space-book}. Indeed, it enables tractable closed-form statistical analysis of pointing-induced fading while preserving a physically meaningful dependence on the receiver optical geometry.

\rev{\subsubsection{Operational FOV ceiling}
The receiver FOV cannot be increased without bound, even though a larger $\theta_\text{fov}$ improves the receiver-side stability parameter $\phi_\text{rx}$ defined in Section~\ref{III-stat}. Two practical mechanisms upper-bound $\theta_\text{fov}$ in operational OISL terminals. First, the background-induced shot-noise variance scales as $\sigma_\text{bg}^2 \propto i_\text{bg} \propto \theta_\text{fov}^2$, so wider FOV admits background photons quadratically \cite{deep-space-book}. For shielded space-based terminals operating outside narrow solar-exclusion angles, this contribution is typically small but is not negligible at very wide FOV. Second, the false-lock probability of the closed-loop tracking system grows monotonically with the angular search space, since spurious acquisition correlation peaks scale with the solid angle searched per loop cycle. The two mechanisms jointly define an operational FOV ceiling $\theta_\text{fov,max}$ that enters the constrained optimization of Section~\ref{sec:asymptotic} as the upper bound on $\theta_\text{fov}$. For the IM/DD OOK reference receiver of Table~\ref{tab:linkbudget} in Section~\ref{sec:numerical}, the dominant noise term is thermal/shot-noise rather than background, and the false-lock-driven ceiling is the binding constraint at $\theta_\text{fov,max}\approx 50\,\mu$rad.}

\subsection{Pointing Error Statistics}
The stochastic nature of the channel is primarily induced by mechanical vibrations, attitude control noise, and tracking inaccuracies of the satellite platforms. These effects result in pointing jitter. In practical OISL terminals, slowly varying bias errors may also arise from thermo-mechanical drift and calibration imperfections in the PAT subsystem. Such bias terms are typically estimated and compensated through periodic calibration or closed-loop tracking, leaving a residual stochastic misalignment around the corrected boresight direction \cite{jsac-channel}. 
The residual pointing error at each terminal is therefore modeled as a two-dimensional independent and identically distributed Gaussian process in orthogonal angular axes (e.g., elevation and azimuth) \cite{pointing-rayleigh}. Consequently, the radial pointing error magnitude, $\theta_i$, follows a Rayleigh distribution \cite{pointing-rayleigh2}, which is widely adopted for modeling zero-mean residual pointing jitter in free-space optical links
\begin{equation}
	\label{pointing-pdf1}
	f_{\Theta_i}(\theta_i) = \frac{\theta_i}{\sigma_i^2} \exp\left( -\frac{\theta_i^2}{2\sigma_i^2} \right), \quad \theta_i \ge 0, \quad i \in \{\text{tx}, \text{rx}\},
\end{equation}
where $\sigma_i$ is the standard deviation of the one-dimensional jitter (the jitter angle) at the respective terminal. As we will discuss in the following sections, the variance $\sigma_i^2$ is a key parameter quantifying platform stability.

\rev{It is worth noting that the physical origins of transmitter and receiver residual pointing errors differ substantially in operational OISL terminals. At the transmitter, the residual budget is dominated by spacecraft attitude disturbances (reaction-wheel jitter, micro-vibrations, thermo-mechanical drift) and by fine-steering-mirror servo error, while at the receiver, it is dominated by tracking-loop residuals (finite loop bandwidth, sensor noise) and by detector-plane coupling effects. The use of a common Gaussian/Rayleigh formulation does not assert that these processes are physically identical. Rather, it captures the statistical equivalence of the residual angular misalignment in the steady-state tracking regime. Indeed, the two terminals are characterized by distinct standard deviations $\sigma_\text{tx}$ and $\sigma_\text{rx}$ that reflect their different physical origins.}

The total instantaneous channel gain is thus the product of the aforementioned components, scaled by lumped optical efficiencies as
\begin{equation}
	\label{eq:Gch}
	G_\text{ch} = \eta_\text{tx} \eta_\text{rx} \eta_{\text{opt}} \cdot G_\text{tx0}\rev{\,\eta_T\,\widetilde{L}_\text{tx}(\theta_\text{tx})} \cdot L_s(z) \cdot G_\text{rx0}\rev{\,\eta_R\,\widetilde{L}_\text{rx}(\theta_\text{rx})},
\end{equation}
where $\eta_\text{tx}$ and $\eta_\text{rx}$ account for transmitter and receiver optical losses (e.g., mirror reflectivity, lens transmission), and $\eta_{\text{opt}}$ encompasses residual end to end optical and opto electronic efficiency factors that are not captured by the aperture gains or the far field beam pattern. For space-vacuum conditions, we also assume that the atmospheric losses are negligible. \rev{Grouping all deterministic terms into a single peak channel gain $G_\text{peak} \triangleq \eta_\text{tx}\eta_\text{rx}\eta_\text{opt}\,\eta_T\eta_R\,G_\text{tx0}\,L_s(z)\,G_\text{rx0}$, the instantaneous channel gain reduces to $G_\text{ch} = G_\text{peak}\,\widetilde{L}_\text{tx}(\theta_\text{tx})\,\widetilde{L}_\text{rx}(\theta_\text{rx})$, in which the stochastic loss factors $\widetilde{L}_\text{tx}$ and $\widetilde{L}_\text{rx}$ satisfy $\widetilde{L}_i(0)=1$ and are statistically characterized in Section~\ref{III-stat}.}

\rev{\subsection{Model Scope and Assumptions}
\label{sec:model_scope}
The canonical Rayleigh-jitter formulation \eqref{pointing-pdf1} adopted throughout this paper rests on four working assumptions that are appropriate for the steady-state closed-loop tracking regime of modern OISL terminals:
\begin{itemize}
	\item[(A1)] {Bias-compensated regime:} deterministic boresight bias has been removed by the PAT subsystem, so that the residual angular misalignment is approximately zero-mean.
	\item[(A2)] {Isotropy:} the two orthogonal angular jitter components (azimuth and elevation) are approximately independent and have approximately equal variance.
	\item[(A3)] {Local stationarity:} the disturbance statistics are stationary over the analysis interval (typically a frame or burst).
	\item[(A4)] {Inter-terminal independence:} the transmitter and receiver jitter processes are independent. This is reasonable because the two terminals reside on physically separate spacecraft. The implications of relaxing (A4) for bidirectional links are discussed in ~\ref{sec:asymptotic}.
\end{itemize}
Under (A1)--(A3), each $\Theta_i$ admits the Rayleigh density \eqref{pointing-pdf1} with terminal-specific scale $\sigma_i$. The scope of (A1)--(A4) is consistent with the operational regime of modern PAT-equipped OISL. Nevertheless, to obtain a deep insight into the proposed model, we will quantify departures from these assumptions in the next subsection.}

\rev{\subsection{Extensions Beyond the Canonical Rayleigh Model}
\label{sec:rayleigh_extensions}
Two operationally relevant generalizations of the canonical Rayleigh model deserve explicit treatment, since they relax assumptions (A1) and (A2) of Section~\ref{sec:model_scope}.

\subsubsection{Anisotropic jitter (Beckmann/Hoyt)} When the orthogonal angular variances $\sigma_x$ and $\sigma_y$ differ, the radial pointing error follows a zero-mean Beckmann (Hoyt) distribution. Defining the anisotropy parameter $q \triangleq \min(\sigma_x,\sigma_y)/\max(\sigma_x,\sigma_y)\in(0,1]$, $q=1$ recovers the Rayleigh model, and for small anisotropy (i.e., when $q\to 1$) the distribution is well-approximated by a Rayleigh distribution with geometric-mean scale $\sigma_\text{eq}=\sqrt{\sigma_x\sigma_y}$. The closed-form expressions of Section~\ref{III-stat} remain accurate to within a fraction of a decibel of the required margin in this regime. For severe anisotropy ($q\lesssim 0.5$), the closed-form expressions should be treated as conservative bounds, with the exact Hoyt PDF substituted numerically.

\subsubsection{Residual boresight bias (Rician)} If a residual boresight bias $\mu_b$ remains after PAT compensation, the radial misalignment follows a Rician distribution with non-centrality $\mu_b$ and scale $\sigma$, characterized by the Rician $K$-factor $K=\mu_b^2/(2\sigma^2)$. For $K\ll 1$ the Rician PDF collapses to Rayleigh, and the closed-form expressions apply unchanged. For $K\gtrsim 1$, the deterministic component of the bias can be absorbed into a degraded peak channel gain $G_\text{peak}'=G_\text{peak}\exp(-G_{p,i}\mu_b^2)$, after which the closed-form outage and capacity expressions apply in their original form (provided $\mu_b\ll\theta_\text{div}$). Beyond this regime, the exact Rician PDF must be substituted numerically.

Throughout the remainder of the paper, the analysis is carried out under the canonical Rayleigh model, which captures the dominant operating regime of modern PAT-equipped OISL terminals. A full numerical sensitivity sweep across $q$ and $K$ is identified as future work.}

\section{Statistical Characterization of the Optical Channel}
\label{III-stat}

The instantaneous channel gain $G_\text{ch}$ is a stochastic quantity governed by the random pointing errors $\theta_\text{tx}$ and $\theta_\text{rx}$. To analyze system performance metrics such as outage probability and ergodic capacity, we derive a closed-form statistical model for $G_\text{ch}$. This is achieved by employing accurate Gaussian approximations for the optical gain profiles, which transform the Rayleigh-distributed angular errors into tractable power-law distributions for the gain factors.

\subsection{Gaussian Approximation of Gain Profiles}
While the exact FFIP \eqref{eq:ffip} and receiver coupling \eqref{eq:coupling} involve Bessel and Airy functions, their main lobes can be accurately approximated by Gaussian profiles for small pointing errors ($\theta_i \ll \theta_{\text{div}}, \theta_{\text{fov}}$). This approximation is common in analytical treatments of pointing loss \cite{pointing-1, channel_model5, pointing-3} and is mathematically essential for deriving closed-form statistics. Here, we adopt this approximation and will verify its validity for typical OISL parameters.

The normalized loss factors due to pointing can be described as
\begin{equation}
	\label{approx_gain_eqs}
	\rev{\widetilde{L}_i(\theta_i)} \approx \exp\left( -G_{p,i} \, \theta_i^2 \right), \quad i \in \{\text{tx}, \text{rx}\}.
\end{equation}
\rev{The pointing gain parameter $G_{p,i}$ is obtained from a curvature match (i.e., a second-order Taylor expansion) of the normalized loss factor at $\theta_i=0$, which is equivalent to matching the $1/e^2$ half-angle of the main lobe of the exact response. The two parameters thus arise from a local analytical approximation rather than from curve fitting.} Here, $G_{p,i}$ is the {pointing gain parameter} that characterizes the equivalent width of the Gaussian profile as
\begin{itemize}
	\item {Transmitter ($G_{p,\text{tx}}$)}: Defined as $2 / \theta_{\text{div}}^2$, where $\theta_{\text{div}}$ is the $1/e^2$ half-angle beam divergence. It shows how fast beam power drops with angular mispointing. For an optimally truncated Gaussian beam, $\theta_{\text{div}}$ is a function of the aperture diameter $D_{tx}$ and wavelength $\lambda$. In particular,  the beam divergence in the far-field is given by
	\begin{equation}
		\theta_{\text{div}} = \frac{2\lambda}{\pi D_{\text{tx}}} \, f_{\text{trunc}},
	\end{equation}
	where $f_{\text{trunc}}$ accounts for the aperture truncation \rev{and obscuration} effect, and can be expressed as a function of the \rev{obscuration ratio $\gamma_o$ (evaluated under the optimum-truncation condition)} as	$f_\text{trunc}(\gamma_o) \approx 1.48 - 2.64\,\gamma_o^{2} + 2.84\,\gamma_o^{3}$,
	as reported in \cite{jsac-channel}.
	\item {Receiver ($G_{p,\text{rx}}$)}: Defined as $2 / \theta_{\text{fov}}^2$, where $\theta_{\text{fov}}$ is the $1/e^2$ half-angle of the receiver's Gaussian sensitivity profile, effectively parameterizing the spatial extent of the FOV (typically determined by the detector size and focal length). It actually shows how fast coupling efficiency, \rev{$\widetilde{L}_\text{rx}(\theta_\text{rx})$}, drops with angular misalignment. \rev{Note that $\theta_\text{fov}$ here is an {equivalent Gaussian-FOV} parameter obtained by curvature-matching the exact Airy-on-detector coupling \eqref{eq:coupling} at zero misalignment, not a literal hard-stop FOV. In particular, $\theta_\text{fov}$ relates to the receiver lens focal length $f$, detector active radius $r_d$, and Airy radius $r_\text{Airy}=1.22\lambda f/D_\text{rx}$ through the curvature of the Airy-coupling response at zero displacement. In the limit $r_d\gg r_\text{Airy}$ (large detector), $\theta_\text{fov}$ approaches the geometric ratio $r_d/f$, recovering the conventional geometric FOV. In the limit $r_d\ll r_\text{Airy}$, $\theta_\text{fov}$ is governed by the diffraction-limited Airy lobe and scales as $\lambda/D_\text{rx}$. The Gaussian-FOV parameterization, therefore, interpolates between these two physical regimes.}
\end{itemize}

\subsubsection{Validation of Gaussian Far-Field and Coupling Approximations}

\begin{figure}[htbp]
	\centering
	\includegraphics[width=1\columnwidth]{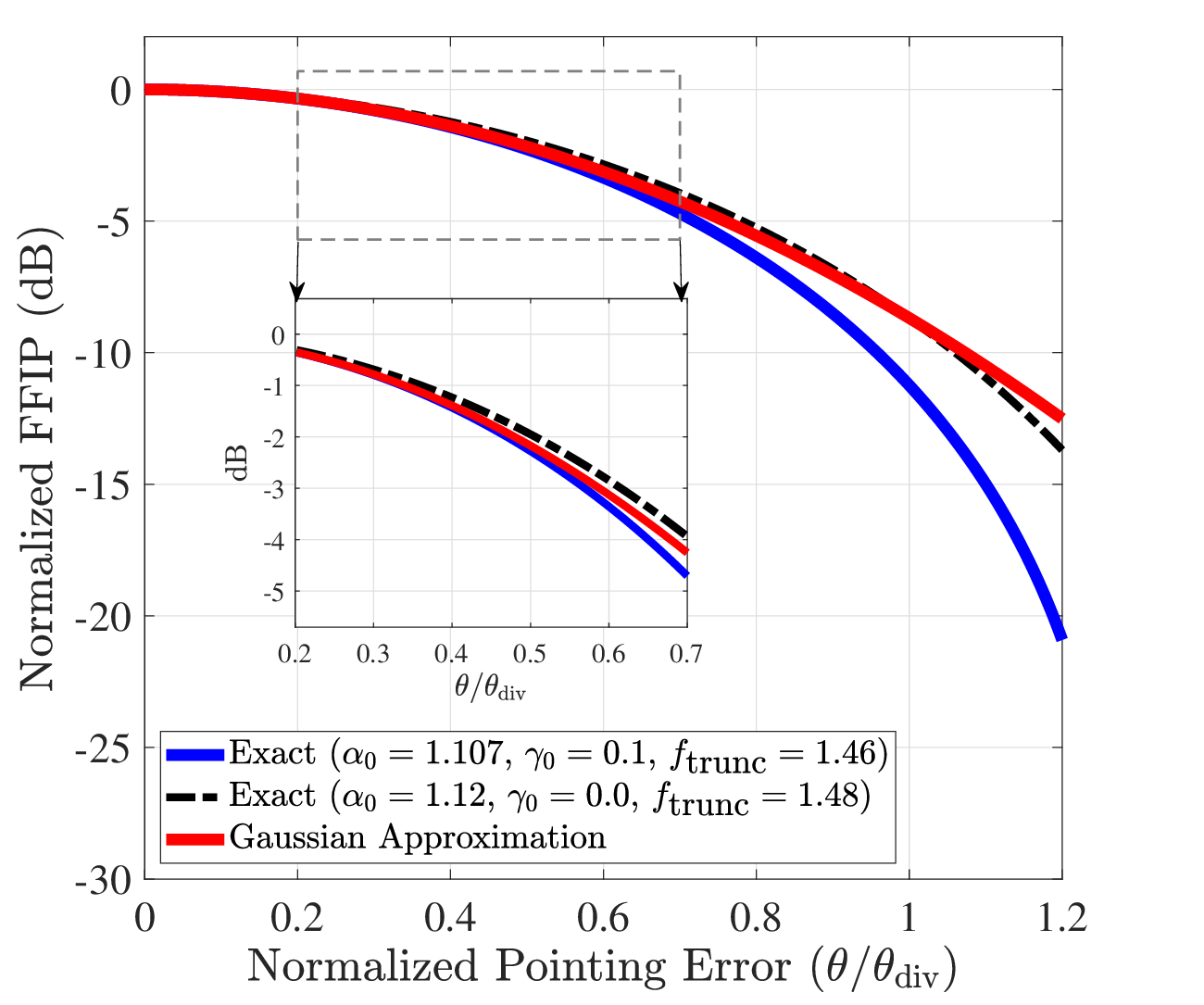}
	\caption{Validation of Gaussian approximation \eqref{approx_gain_eqs} against exact truncated Gaussian beam pattern \eqref{eq:ffip} (deviation remains below 0.5 dB for $\theta/\theta_{\text{div}} < 0.7$).}  
	\label{fig:ffip_validation_tx_side}
\end{figure}

\begin{figure}[htbp]
	\centering
	\includegraphics[width=0.95\columnwidth]{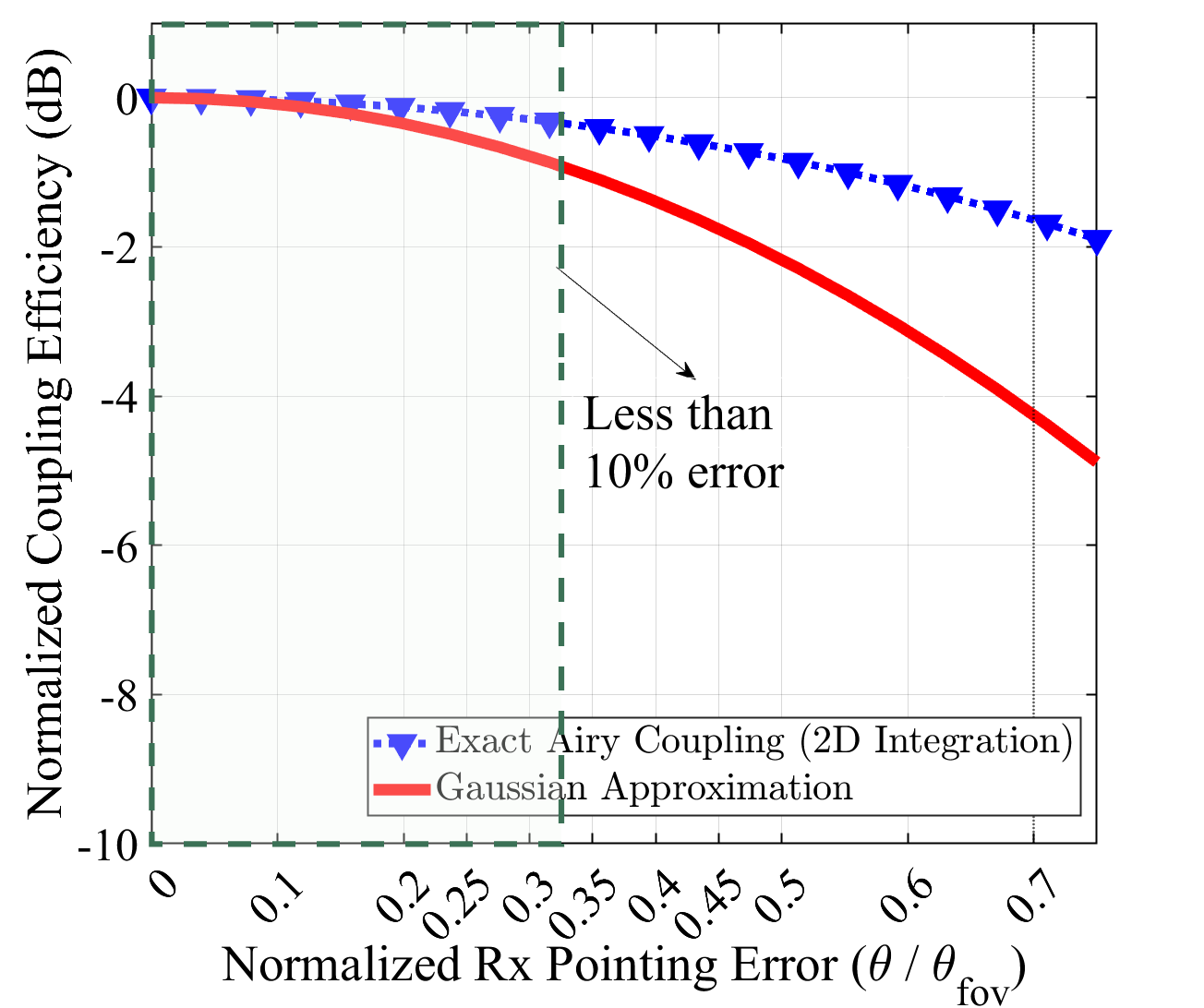}
	\caption{Validation of Gaussian approximation \eqref{approx_gain_eqs} against exact coupling efficiency \eqref{eq:coupling} at the receiver side.}  
\label{fig:ffip_validation_rx_side}
\end{figure}

\rev{
\begin{figure}[htbp]
	\centering
	\includegraphics[width=0.98\columnwidth]{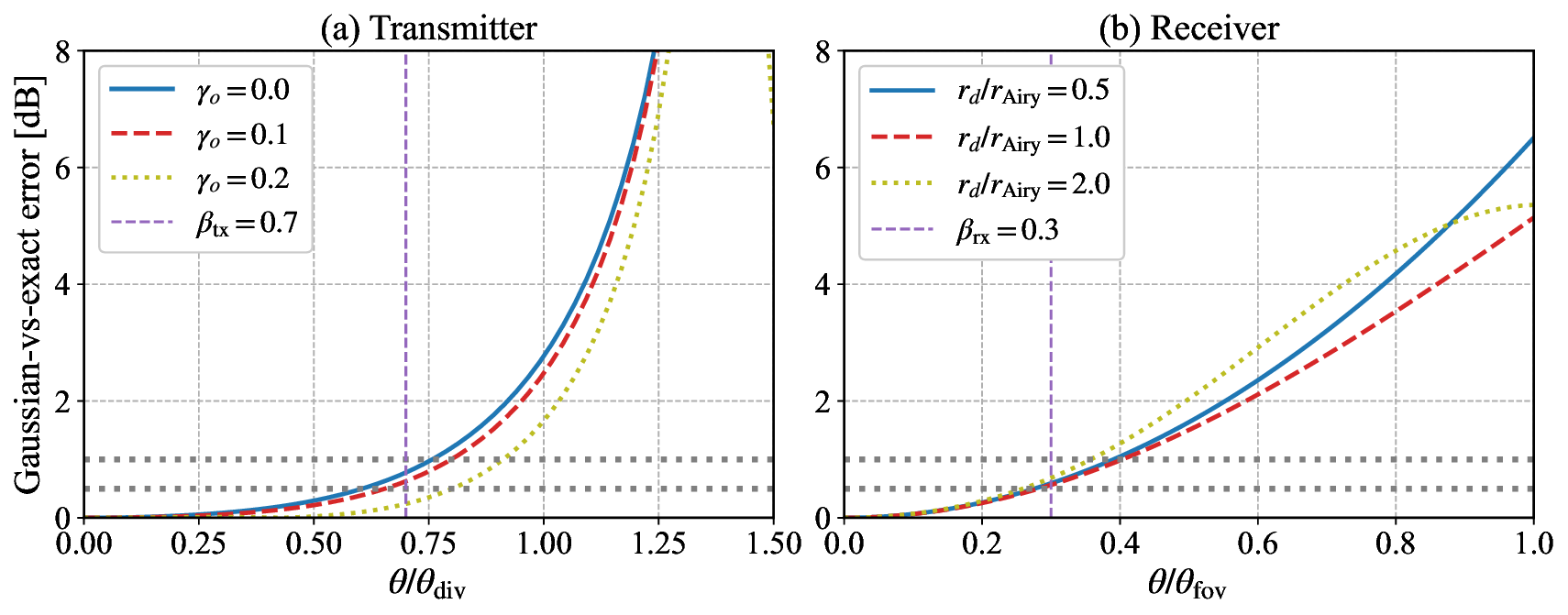}
	\caption{\rev{Absolute approximation error in dB of the Gaussian model against the exact transmitter Bessel-integral response \eqref{eq:ffip} and the exact receiver Airy-on-detector coupling \eqref{eq:coupling}, plotted versus normalized pointing angle. Horizontal reference lines mark $0.5$~dB and $1.0$~dB; vertical dashed lines mark the demonstrated validity radii $\beta_\text{tx}\approx 0.7$ and $\beta_\text{rx}\approx 0.3$.}}
	\label{fig:gauss_error_db}
\end{figure}
}

Fig.~\ref{fig:ffip_validation_tx_side} evaluates the accuracy of the Gaussian approximation against the exact Bessel-integral solution for modelling the FFIP. Two transmitter configurations are examined, i.e., an unobscured aperture ($\gamma_o = 0$, $\alpha_{0} = 1.12$, $f_{\text{trunc}} = 1.48$) and a centrally obscured aperture ($\gamma_o = 0.1$, $\alpha_{0} = 1.107$, $f_{\text{trunc}} = 1.46$). As observed, the Gaussian approximation closely matches the exact solution within the normalized pointing error region $\theta / \theta_{\text{div}} < 0.7$ \rev{(which we refer to as the validity radius $\beta_{\text{tx}}$)}, where the deviation remains below 0.5~dB for both configurations. Notably, in the unobscured case ($\gamma_o = 0$), the approximation retains good accuracy even when the pointing error approaches or slightly exceeds the beam divergence.

This operating regime corresponds to the pointing precision typically achieved by modern PAT systems in OISLs, where closed-loop tracking suppresses residual angular errors to a small fraction of the beam divergence under nominal operating conditions ~\cite{deep-space-book}. Consequently, the dominant contribution to power loss arises from the main-lobe roll-off of the far-field beam pattern, which is accurately captured by the Gaussian approximation. This justifies the use of the Gaussian model for performance evaluation, link margin analysis, and beamwidth optimization in operational OISL scenarios where the link has been successfully acquired and is maintained under closed-loop tracking.
For larger pointing errors, however, the approximation gradually diverges from the exact formulation. At $\theta / \theta_{\text{div}} = 1.0$ with central obscuration ($\gamma_o \neq 0$), the Gaussian model underestimates the received intensity by approximately 2--3~dB, increasing to about 5--6~dB at $\theta / \theta_{\text{div}} = 1.2$. These discrepancies arise from diffraction effects and truncation-induced sidelobes captured by the Bessel-integral formulation but inherently neglected in the Gaussian representation.

Given its sub-0.5~dB accuracy within the nominal PAT operating range ($\theta / \theta_{\text{div}} < 0.7$), the Gaussian approximation is sufficiently accurate for steady-state OISL design tasks such as link margin evaluation, sensitivity analysis, and pointing control optimization, as considered in this paper. During acquisition phases, however, where large misalignment errors may occur ($\theta / \theta_{\text{div}} > 1.0$), exclusive reliance on the Gaussian model may lead to optimistic link margin predictions and underestimation of required transmit power. In such regimes, the exact formulation in~\eqref{eq:ffip} should be employed to ensure conservative and reliable system design.

Furthermore, Fig.~\ref{fig:ffip_validation_rx_side} compares the Gaussian approximation with the exact receiver-side coupling efficiency obtained via numerical two-dimensional integration of the Airy diffraction pattern in~\eqref{eq:coupling} over a circular detector aperture. Results are expressed as a function of normalized pointing error, where the Gaussian model is parameterized to match the $1/e^{2}$ intensity roll-off.
As expected, the Gaussian model does not perfectly reproduce the exact Airy-based coupling profile. However, within the primary tracking regime the approximation captures the dominant decay trend of the main lobe while maintaining analytical simplicity. In particular, for normalized errors below $0.3$ (vertical dashed line--\rev{which we refer to as the validity radius $\beta_{\text{rx}}$)}, corresponding to pointing errors less than $30\%$ of the receiver FOV (typical for OISL links operating under closed-loop tracking \cite{deep-space-book}), the absolute difference in coupling efficiency remains small and translates to a negligible loss difference in practical link-budget calculations.
More accurate numerical representations of the Airy coupling profile could be employed, but they would generally prevent the derivation of closed-form statistical expressions for the channel distribution and outage probability. The Gaussian approximation therefore provides a practical compromise between physical realism and analytical tractability for system-level performance analysis.

\rev{To make these observations quantitatively precise across the obscuration and detector-geometry families relevant to operational OISL design, Fig.~\ref{fig:gauss_error_db} reports the absolute Gaussian-versus-exact error of both responses on a common axis. On the transmitter side (left panel), the error at the validity radius $\beta_\text{tx}=0.7$ remains below approximately $1$~dB across the full obscuration range $\gamma_o\in[0,0.2]$, with the unobscured case $(\gamma_o=0)$ representing the worst case at this radius ($\approx 0.8$~dB) and centrally obscured configurations showing strictly smaller errors. Beyond $\theta/\theta_\text{div}\approx 1$, the error grows rapidly as diffraction sidelobes and aperture-truncation contributions become significant. On the receiver side (right panel), the error at $\beta_\text{rx}=0.3$ is approximately $0.5$~dB across the detector-to-Airy radius range $r_d/r_\text{Airy}\in[0.5,2.0]$, and the three curves are nearly indistinguishable in the small-misalignment regime where the Gaussian curvature-match dominates. The detector-geometry dependence becomes visible only at $\theta/\theta_\text{fov}\gtrsim 0.5$, where residual Airy-ring contributions start to differentiate the three cases. The validity radii $\beta_\text{tx}=0.7$ and $\beta_\text{rx}=0.3$ used in the tail-probability bound \eqref{eq:prob_invalid} of Section~\ref{sec:numerical} are read directly from these curves.}

These results confirm that the approximations in~\eqref{approx_gain_eqs} provide an accurate and computationally efficient framework for characterizing pointing-induced fading in OISL scenarios under nominal tracking conditions. Having evaluated these approximations, we now proceed to derive closed-form expressions for the overall channel gain.

\subsection{Probability Density Function of Pointing-Induced Loss}
As established in \eqref{pointing-pdf1}, the radial pointing error $\Theta$ follows a Rayleigh distribution governed by the jitter standard deviation $\sigma$. By applying the change of variables \rev{$\widetilde{L}_i = \exp(-G_{p,i} \Theta_i^2)$} to the Rayleigh PDF, the PDF of the \rev{normalized} loss factor \rev{$\widetilde{L}_i$} is derived as a power-law distribution, i.e., we have
\begin{equation}
	\label{eq:fLi}
	f_{\rev{\widetilde{L}_i}}(l_i) = \phi_i \, l_i^{\phi_i - 1}, \quad 0 \leq l_i \leq 1, \quad i \in \{\text{tx}, \text{rx}\},
\end{equation}
where $\phi_i$ is the dimensionless {stability parameter}. This distribution has a heavy tail toward zero when $\phi_i < 1$, indicating a high probability of deep fades in poorly stabilized links, while for $\phi_i > 1$ the density concentrates near $l_i = 1$, corresponding to stable links with infrequent severe misalignment.  

\rev{To record the explicit derivation of \eqref{eq:fLi}, from $\widetilde{L}_i = \exp(-G_{p,i}\Theta_i^2)$, we obtain $\Theta_i=\sqrt{-\ln \widetilde{L}_i / G_{p,i}}$ and the Jacobian $|d\theta_i/dl_i| = (2 l_i\,G_{p,i}\theta_i)^{-1}$. Substituting into \eqref{pointing-pdf1} gives
\begin{equation}
	f_{\widetilde{L}_i}(l_i) = f_{\Theta_i}\bigl(\theta_i(l_i)\bigr)\left|\frac{d\theta_i}{dl_i}\right| = \frac{1}{2 G_{p,i}\sigma_i^2}\,l_i^{\,1/(2G_{p,i}\sigma_i^2)\,-\,1},
\end{equation}
which recovers \eqref{eq:fLi} with $\phi_i \triangleq 1/(2 G_{p,i}\sigma_i^2)$.}

\rev{With $G_{p,\text{tx}}=2/\theta_\text{div}^2$ and $G_{p,\text{rx}}=2/\theta_\text{fov}^2$,} we define the specific parameters for the link terminals as
\begin{equation}
	\label{stability_parameters}
	\phi_\text{tx} = \frac{\theta_\text{div}^2}{4\sigma_\text{tx}^2}, \quad \text{and} \quad \phi_\text{rx} = \frac{\theta_\text{fov}^2}{4\sigma_\text{rx}^2}.
\end{equation}
The parameter $\phi_i$ therefore provides a compact and physically interpretable measure of terminal stability. Larger values of $\phi_i$ arise from wider beam divergence or receiver FOV, and from improved pointing stability (smaller $\sigma_i$). In contrast, narrow beams combined with large platform jitter lead to small $\phi_i$, which results in a heavy-tailed loss distribution and frequent deep fades. This formulation provides a direct quantitative link between optical design parameters, platform stability, and the statistical severity of pointing-induced fading, and will be central to the outage and asymptotic analyses developed in the sequel.


\subsection{Composite PDF of the Total Channel Gain}
To obtain the end-to-end channel PDF, one can define the total channel gain as
\begin{equation}
	G_\text{ch} = G_{\text{peak}} \cdot \rev{\widetilde{L}_\text{tx} \cdot \widetilde{L}_\text{rx}},
\end{equation}
 where \rev{$G_{\text{peak}} = \eta_\text{tx}\eta_\text{rx}\eta_{\text{opt}}\,\eta_T\eta_R\, G_{tx0} L_s G_{rx0}$} encompasses all deterministic gains and efficiencies, \rev{including the truncation/obscuration efficiency $\eta_T$ at the transmitter and the on-axis Airy-to-detector coupling efficiency $\eta_R$ at the receiver introduced in Section~\ref{sec:sysmodel}.} Let $Z = \rev{\widetilde{L}_\text{tx} \widetilde{L}_\text{rx}}$ be the product of two independent power-law random variables with parameters $\phi_\text{tx}$ and $\phi_\text{rx}$, respectively. \rev{For mathematical transparency, $f_Z(z)$ can equivalently be obtained by Mellin-transform methods \cite{Mellin-transform}. Accordingly, the Mellin transform of \eqref{eq:fLi} is $\mathcal{M}[f_{\widetilde{L}_i}](s) = \phi_i/(\phi_i + s - 1)$, and the Mellin transform of the product of independent variables factors as $\mathcal{M}[f_Z](s) = \phi_\text{tx}\phi_\text{rx}/[(\phi_\text{tx}+s-1)(\phi_\text{rx}+s-1)]$. Subsequently, partial-fraction inversion recovers \eqref{int-pdf} below.} The PDF of $Z$ can thus be found via the product convolution integral (assuming $\phi_\text{tx} \neq \phi_\text{rx}$) as
\begin{equation}
	\label{int-pdf}
	\begin{split}
		f_Z(z) &= \int_z^1 f_{\widetilde{L}_\text{tx}}(x) f_{\widetilde{L}_\text{rx}}(z/x) \frac{1}{x} \, dx \\
		&= \frac{\phi_\text{tx} \phi_\text{rx}}{\phi_\text{tx} - \phi_\text{rx}} \left( z^{\phi_\text{rx}-1} - z^{\phi_\text{tx}-1} \right), \quad 0 \leq z \leq 1.
	\end{split}
\end{equation}
Applying the scaling transformation $g = G_{\text{peak}} z$, we can finally write the closed-form PDF for the channel gain as
\begin{equation}
	\label{main-pdf}
	\begin{split}
		f_{G_\text{ch}}(g) &= \frac{\phi_\text{tx} \phi_\text{rx}}{G_{\text{peak}}(\phi_\text{tx} - \phi_\text{rx})} \\
		&\quad \times \left[ \left(\frac{g}{G_{\text{peak}}}\right)^{\phi_\text{rx}-1} - \left(\frac{g}{G_{\text{peak}}}\right)^{\phi_\text{tx}-1} \right], \\
		& \qquad 0 < g \leq G_{\text{peak}}.
	\end{split}
\end{equation}
Note that this PDF may diverge at $g=0$ if $\phi_\text{tx}<1$ or $\phi_\text{rx}<1$, which corresponds to a high probability of very deep fades in unstable links. 

\subsubsection{Special Case (Symmetric Stability Parameters)}
When the transmitter and receiver have identical stability parameters, i.e., $\phi_\text{tx} = \phi_\text{rx} = \phi$, the integral in \eqref{int-pdf} must be evaluated separately. \rev{An equivalent and more transparent derivation proceeds by the substitution $U_i \triangleq -\ln\widetilde{L}_i$, which converts each power-law variable to an exponential $U_i \sim \mathrm{Exp}(\phi)$. The product $Z = \widetilde{L}_\text{tx}\widetilde{L}_\text{rx}$ then corresponds to $V \triangleq U_\text{tx} + U_\text{rx}$, which is a sum of two i.i.d.\ exponentials of rate $\phi$, hence Erlang-2 distributed with density $f_V(v)=\phi^2 v\,e^{-\phi v}$ for $v\ge 0$. Transforming back via $Z = e^{-V}$ gives the result below, where the $-\ln z$ factor is the image of the linear-in-$v$ Erlang-2 factor under the inverse exponential map.} Performing the convolution yields
\begin{equation}
	f_Z(z) = -\phi^2 \, z^{\phi-1} \ln(z), \quad 0 < z \leq 1.
\end{equation}
Thus, the corresponding PDF for the channel gain is 
\begin{equation}
	\label{special-pdf}
	{\small
	f_{G_\text{ch}}(g) = -\frac{\phi^2}{G_{\text{peak}}} \left( \frac{g}{G_{\text{peak}}} \right)^{\phi-1} \ln\!\left( \frac{g}{G_{\text{peak}}} \right), \quad 0 < g \leq G_{\text{peak}}}.
\end{equation}

\section{Outage Probability Analysis}
\label{sec:outage}

A key performance metric in OISL design is the outage probability, $P_{\text{out}}$, defined as the probability that the instantaneous received signal-to-noise ratio (SNR) falls below a target threshold $\gamma_{\text{th}}$ required to sustain a desired communication quality, such as a target bit-error ratio. Formally, the outage probability is expressed as
\begin{equation}
	P_{\text{out}} = \Pr\!\left( \gamma < \gamma_{\text{th}} \right),
\end{equation}
where $\gamma$ denotes the instantaneous electrical SNR at the receiver.
Since the received SNR is proportional to the instantaneous optical channel gain, the outage event can be equivalently expressed in terms of the channel gain as
\begin{equation}
	P_{\text{out}} 
	= \Pr\!\left( G_{\text{ch}} < G_{\text{th}} \right),
\end{equation}
where $G_{\text{th}}$ is the minimum channel gain required to meet the SNR threshold $\gamma_{\text{th}}$, and $\Pr(\cdot)$ denotes probability. In the following section, we derive a closed-form expression for this metric and discuss the resulting design insights. These insights illustrate how outage analysis can guide the optimization of OISL parameters and enhance overall link reliability.

\subsection{Closed-Form Outage Probability Derivation}
The outage probability is obtained by integrating the composite PDF in \eqref{main-pdf} from $0$ to $G_{\text{th}}$, i.e.,
\begin{align}
	\label{outage_simp}
	P_{\text{out}}(G_{\text{th}}) &= \int_0^{G_{\text{th}}} f_{G_\text{ch}}(g) \, dg \nonumber \\
	&= \frac{1}{\phi_\text{tx} - \phi_\text{rx}} \left[ \phi_\text{tx} \left(\frac{G_{\text{th}}}{G_{\text{peak}}}\right)^{\phi_\text{rx}} - \phi_\text{rx} \left(\frac{G_{\text{th}}}{G_{\text{peak}}}\right)^{\phi_\text{tx}} \right].
\end{align}
Also, integrating the PDF in \eqref{special-pdf} gives the outage probability for the symmetric case as
\begin{equation}
	\label{outage_symetric}
	P_{\text{out}}(G_{\text{th}}) = \left( \frac{G_{\text{th}}}{G_{\text{peak}}} \right)^{\phi} \left[ 1 - \phi \ln\!\left( \frac{G_{\text{th}}}{G_{\text{peak}}} \right) \right].
\end{equation}

\subsection{Design Interpretation and Tradeoffs}
The ratio $G_{\text{th}} / G_{\text{peak}}$ represents the inverse of the link margin, $M = G_{\text{peak}} / G_{\text{th}}$. 
This relationship highlights an important tradeoff between link margin and pointing stability. 

First, increasing $M$, improves the outage performance, although the improvement rate is limited by the pointing stability parameters $\phi_{\text{tx}}$ and $\phi_{\text{rx}}$. When the stability is poor, meaning that the corresponding $\phi$ is small, the outage curve becomes shallow and large power increases are required to obtain even modest reductions in $P_{\text{out}}$. We will investigate this problem through the asymptotic analysis presented in Section~\ref{sec:asymptotic}.

A second important implication concerns beamwidth optimization. While the existence of a tradeoff between geometric loss and pointing robustness has been recognized in prior free-space optical and inter-satellite link studies, our framework places this tradeoff within a unified end-to-end statistical model that explicitly accounts for independent pointing jitter at both terminals. Since the stability parameter satisfies $\phi_{\text{tx}} \propto \theta_{\text{div}}^{2} / \sigma_{\text{tx}}^{2}$, the beam divergence directly controls not only the average link budget but also the asymptotic outage behavior through the weakest-link principle derived in Section~\ref{sec:asymptotic}. This reveals that beamwidth selection affects the outage slope when the transmitter is the bottleneck terminal, rather than merely shifting the outage curve. Moreover, the resulting optimum must be interpreted under the physical constraint that $\theta_{\text{div}}$ is coupled to the transmit aperture size, which introduces nontrivial payload and acquisition tradeoffs. These aspects are examined quantitatively in Section~\ref{sec:numerical}.

\section{Asymptotic Performance Analysis}
\label{sec:asymptotic}

Building on the closed-form outage expression derived in Section~\ref{sec:outage}, we now examine the high-margin regime to extract asymptotic performance metrics and obtain sharper design insights. 

\subsection{Asymptotic Outage Behaviour}
Recall that the link margin was defined as $M = G_{\text{peak}}/G_{\text{th}}$. Substituting this definition into \eqref{outage_simp}, the outage probability can be written explicitly as a function of the link margin, i.e, we have
\begin{equation}
	\label{eq:outage_margin}
	P_{\text{out}}(M) = \frac{1}{\phi_\text{tx} - \phi_\text{rx}}
	\left[
	\phi_\text{tx} M^{-\phi_\text{rx}} - \phi_\text{rx} M^{-\phi_\text{tx}}
	\right], \quad M \geq 1.
\end{equation}
This representation makes explicit how the outage probability decays as a power law of the inverse margin, as discussed in the design interpretation of Section~\ref{sec:outage}.

\subsubsection{Decay Exponent}

In the high-margin regime ($M \gg 1$), the decay rate of $P_{\text{out}}$ is governed by the smaller of the two stability parameters. Defining
\[
\phi_{\min} = \min(\phi_\text{tx}, \phi_\text{rx}), 
\qquad
\phi_{\max} = \max(\phi_\text{tx}, \phi_\text{rx}),
\]
and noting that \rev{$M^{-\phi_{\max}}/M^{-\phi_{\min}} = M^{-(\phi_{\max}-\phi_{\min})} \to 0$ as $M\to\infty$ whenever $\phi_\text{tx}\ne\phi_\text{rx}$, so that $M^{-\phi_{\max}} \ll M^{-\phi_{\min}}$ for $M\to\infty$}, the outage probability in \eqref{eq:outage_margin} admits the asymptotic approximation below
\begin{equation}
	\label{eq:outage_asymptotic}
	P_{\text{out}}(M) \sim 
	\frac{\phi_{\max}}{|\phi_\text{tx} - \phi_\text{rx}|}
	M^{-\phi_{\min}}, 
	\qquad M \to \infty.
\end{equation}
This expression shows that the outage probability decays polynomially with margin. \footnote{\rev{The relative error of \eqref{eq:outage_asymptotic} with respect to the exact expression \eqref{eq:outage_margin} scales as $M^{-(\phi_{\max}-\phi_{\min})}$, which vanishes slowly when the two stabilities are nearly balanced; the symmetric form \eqref{asymp_symetric} should be used in that regime.}}

Let us define the asymptotic decay exponent as the asymptotic slope of the outage curve on a log-log scale. Thus we have
\begin{equation}
	\label{eq:diversity_def}
	d \triangleq -\lim_{M \to \infty} 
	\frac{\log P_{\text{out}}(M)}{\log M}.
\end{equation}
Substituting \eqref{eq:outage_asymptotic} into \eqref{eq:diversity_def} yields 
\begin{equation}
	\label{eq:diversity_result}
	d = \phi_{\min}.
\end{equation}
Hence, the asymptotic decay exponent is determined solely by the weaker of the two terminal stabilities. 

\subsubsection{Power Offset}
\label{subsec:coding_gain}
While the asymptotic decay exponent determines the slope of the outage probability curve in the high-margin regime, the {asymptotic power offset} determines its horizontal shift. From \eqref{eq:outage_asymptotic}, the power offset $G_c$ is defined through
\begin{equation}
	P_{\text{out}}(M) \sim (G_c M)^{-\phi_{\min}}, \qquad M \to \infty,
\end{equation}
which yields
\begin{equation}
	\label{eq:coding_gain}
	G_c = \left( \frac{\phi_{\max}}{|\phi_\text{tx} - \phi_\text{rx}|} \right)^{-1/\phi_{\min}}.
\end{equation}
A larger $G_c$ corresponds to a lower outage probability at a given margin. The power offset captures the imbalance between the terminal stabilities and does not affect the asymptotic decay exponent $d = \phi_{\min}$. When the weaker terminal has low stability (small $\phi_{\min}$), the exponent $1/\phi_{\min}$ amplifies the impact of imbalance, making the absolute outage performance highly sensitive to differences between $\phi_\text{tx}$ and $\phi_\text{rx}$. In the highly imbalanced regime ($\phi_{\max} \gg \phi_{\min}$), $\phi_{\max}/|\phi_\text{tx} - \phi_\text{rx}| \approx 1$, and thus $G_c \approx 1$, indicating that the weakest terminal dominates the absolute performance.

When the terminal stabilities are nearly symmetric ($\phi_\text{tx} \approx \phi_\text{rx}$), \eqref{eq:coding_gain} becomes ill-conditioned due to the vanishing denominator.  In this regime, the exact symmetric outage expression in \eqref{outage_symetric} should be used. For large margin $M$, we have the symmetric outage written in margin form as
\begin{equation}
	\label{asymp_symetric}
	P_{\text{out}}(M) = M^{-\phi}\bigl(1 + \phi \ln M\bigr), 
	\qquad M \to \infty,
\end{equation}
with $\phi = \phi_\text{tx} = \phi_\text{rx}$. Again, \eqref{asymp_symetric} confirms that the asymptotic decay exponent remains $d = \phi$, with an additional logarithmic offset.

\subsubsection{Design Implications (Weakest-Link Principle)}
\label{subsec:weakest_link}
\rev{The asymptotic analysis above reveals a \emph{high-margin} weakest-link principle in OISL design.} That is, \rev{in the regime $M\to\infty$,} the terminal with poorer normalized pointing stability governs the asymptotic reliability of the link. Specifically, the asymptotic decay exponent is given by $d = \phi_{\min}$. In contrast, improvements to the more stable terminal affect only the asymptotic power offset $G_c$ and do not change the fundamental slope of the outage curve. \rev{Note that this is an asymptotic statement. That is, at moderate margins or when the two stabilities are nearly balanced, the polynomial slope may not yet be visible on a log--log plot, and the exact expressions \eqref{outage_simp} or \eqref{outage_symetric} should be used. The principle is therefore most useful as a design guideline in the high-availability regime ($P_\text{out}\lesssim 10^{-3}$), which is the operational regime of interest for high-reliability OISLs.}

When one terminal exhibits high stability ($\phi_{\max} \gg 1$) and the other suffers from poor tracking ($\phi_{\min} \approx 1$), the outage probability decays slowly with increasing margin. In this regime, increasing transmit power yields only limited improvements in reliability, making power scaling an inefficient means of enhancing availability. Conversely, further enhancing the already stable terminal increases $\phi_{\max}$ and improves the power offset in \eqref{eq:coding_gain}, resulting in a constant horizontal shift of the outage curve without altering its asymptotic decay exponent.

Substantial improvements in link availability therefore require increasing $\phi_{\min}$. This can be achieved by reducing the platform jitter $\sigma$ at the less stable terminal through improved attitude control, vibration isolation, or tracking algorithms, and by optimizing the beam divergence or receiver FOV at that terminal to increase the effective angular tolerance, at the expense of increased geometric loss or background noise. For a target asymptotic decay exponent $d_{\text{target}}$, the system must satisfy $\min(\phi_\text{tx}, \phi_\text{rx}) \geq d_{\text{target}}$, which provides explicit specifications for the pointing and optical subsystems.

\subsection{\rev{High-SNR Spectral-Efficiency Penalty}}
\label{subsec:capacity}
Beyond outage probability, the ergodic capacity provides a measure of the long-term achievable spectral efficiency. The instantaneous spectral efficiency of the OISL channel is given by
\begin{equation}
	C = \log_2\!\left( 1 + \gamma G_{\text{ch}}^\xi \right),
    \label{capacity-1}
\end{equation}
where $\gamma$ is the reference electrical SNR, $G_{\text{ch}}$ is the normalized optical channel gain, and $\xi$ is the detection coefficient. For coherent detection, $\xi = 1$, whereas for standard intensity modulation and direct detection (IM/DD), $\xi = 2$ due to the square-law nature of photodetectors. \rev{The exact finite-SNR ergodic capacity is the expectation of $C$ over the channel-gain distribution, which can be expressed as a single one-dimensional integral over the closed-form PDF \eqref{main-pdf} as
\begin{equation}
	\label{eq:exact_ergodic_capacity}
	\bar{C}(\gamma) = \int_0^{G_\text{peak}} \log_2\bigl(1+\gamma g^\xi\bigr)\,f_{G_\text{ch}}(g)\,dg,
\end{equation}
which can be evaluated numerically. For coherent detection $(\xi = 1)$, \ref{eq:exact_ergodic_capacity} is the exact finite-SNR ergodic capacity \footnote{\rev{For IM/DD $(\xi = 2)$, it is the expectation of the spectral-efficiency proxy $\log_2(1 + \gamma G_{\mathrm{ch}}^2)$ over the gain distribution and inherits the same heuristic nature as \ref{capacity-1}.}}.}

\rev{Furthermore, to extract a tractable closed-form design insight,} in the high-SNR regime ($\gamma \gg 1$), the capacity can be approximated as
\begin{equation}
	C \approx \log_2(\gamma G_{\text{ch}}^\xi) 
	= \log_2(\gamma) + \xi \log_2(G_{\text{ch}}).
\end{equation}
The first term represents the capacity of a static channel, while the second term captures the ergodic capacity penalty induced by pointing jitter \rev{in the high-SNR slope}.

Using the derived PDF of the loss factors \rev{$f_{\widetilde{L}_i}(l_i) = \phi_i l_i^{\phi_i - 1}$}, the expected logarithmic loss for terminal $i$ is
\begin{align}
	\mathbb{E}[\ln(\rev{\widetilde{L}_i})] 
	&= \int_0^1 \ln(x) \, \phi_i x^{\phi_i-1} \, dx 
	= -\frac{1}{\phi_i}\rev{,} \quad i \in \{\text{tx}, \text{rx}\}.
\end{align}
Converting to base-2 logarithms and summing the transmitter and receiver contributions yields the closed-form \rev{high-SNR ergodic capacity penalty}
\begin{equation}
	\label{eq:capacity_penalty}
	\Delta C_{\text{jitter}} 
	\approx - \frac{\xi}{\ln 2} 
	\left( \frac{1}{\phi_{\text{tx}}} + \frac{1}{\phi_{\text{rx}}} \right)
	\quad \text{[bits/s/Hz]}.
\end{equation}
\rev{Here, \eqref{eq:capacity_penalty} is the high-SNR slope of $\bar{C}(\gamma)$ from \eqref{eq:exact_ergodic_capacity}, not an exact finite-SNR penalty.  At moderate SNR, \eqref{eq:exact_ergodic_capacity} should be used directly.}

This result provides a complementary design insight to the outage analysis. While the outage probability is dominated by the weakest terminal through $\min(\phi_{\text{tx}}, \phi_{\text{rx}})$, the ergodic capacity degradation depends on the sum of the inverse stability parameters of both terminals. 

\rev{\subsubsection{IM/DD Scope}
\label{subsubsec:imdd_scope}
The expression $\log_2(1+\gamma G_\text{ch}^2)$ is adopted as a {heuristic spectral-efficiency proxy} for direct-detection systems, in the same spirit as the standard coherent capacity model. It captures the squared-channel dependence of IM/DD receivers but does not constitute the information-theoretic IM/DD capacity, which is shaped by nonnegative-input and average/peak optical-power constraints, and by shot, thermal, and background noise contributions whose statistics depend on the detector architecture. For the IM/DD OOK reference receiver of Table~\ref{tab:linkbudget}, the SNR threshold $\gamma_\text{th}$ is obtained from the standard Gaussian-approximation BER analysis of OOK under shot, thermal, and (negligible) background noise; the resulting threshold mapping is used in Section~\ref{sec:numerical} only for link-margin calibration, not as an information-theoretic capacity claim.}

\rev{\subsection{Constrained Beamwidth Optimization}
\label{subsec:opt_problem}
The design insights of Sections~\ref{subsec:weakest_link} and \ref{subsec:capacity} can be made operationally precise by stating the underlying optimization problem explicitly. With $\theta_\text{div}$, $\theta_\text{fov}$, and $P_\text{tx}$ as the three principal design variables, the constrained OISL design problem may be written as
\begin{equation}
	\label{eq:P1}
	\textrm{(P1):}\quad \min_{\theta_\text{div},\,\theta_\text{fov},\,P_\text{tx}} \; P_\text{out}(\theta_\text{div},\theta_\text{fov},P_\text{tx}),
\end{equation}
subject to
\begin{align*}
	&\theta_\text{div}\ge\theta_\text{div,min},\quad \theta_\text{fov}\le\theta_\text{fov,max},\\
	&0<P_\text{tx}\le P_\text{tx,max},\quad D_\text{tx}(\theta_\text{div})\le D_\text{tx,max}.
\end{align*}
The dependence of the peak channel gain on the design variables is made explicit through the optimally-truncated Klein--Degnan relation $D_\text{tx}(\theta_\text{div})=(2 f_\text{trunc}(\gamma_o)/\pi)(\lambda/\theta_\text{div})$, which gives the peak transmitter aperture-gain scaling $G_\text{tx0}\propto\theta_\text{div}^{-2}$. The objective \eqref{eq:P1} is evaluated using \eqref{outage_simp} (or \eqref{outage_symetric} when $\phi_\text{tx}=\phi_\text{rx}$) with the stability parameters \eqref{stability_parameters}. The FOV upper bound $\theta_\text{fov,max}$ is the operational ceiling defined in Section~\ref{sec:sysmodel}-C.

\subsubsection{First-order conditions} At an interior optimum of Problem~\eqref{eq:P1} in the transmitter-limited regime ($\phi_\text{tx}<\phi_\text{rx}$), $P_\text{tx}=P_\text{tx,max}$ is active and the stationary condition with respect to $\theta_\text{div}$ reduces to the transcendental relation
\begin{equation}
	\label{eq:kkt_thetadiv}
	\theta_\text{div}^{\star\,2} \;=\; 4\sigma_\text{tx}^2\,\ln M\bigl(\theta_\text{div}^\star,P_\text{tx,max}\bigr),
\end{equation}
which gives the intuitive scaling $\theta_\text{div}^\star \approx 2\sigma_\text{tx}\sqrt{\ln M}$ and is consistent with the empirical optimum visible in Fig.~\ref{fig:beamwidth_optimization}. When the optimum predicted by \eqref{eq:kkt_thetadiv} falls below $\theta_\text{div,min}$, the aperture-floor constraint becomes active. The balanced-stability rule $\theta_{\text{div},A}/\theta_{\text{div},B}=\sigma_A/\sigma_B$ for bidirectional asymmetric terminals (introduced in ~\ref{sec:asymptotic}) is the simultaneous-stationarity condition of \eqref{eq:P1} applied at both directions, rather than an asserted optimum.

We provide Algorithm~\ref{design:alg} in Section~\ref{sec:numerical} as the coordinate-descent solver for Problem~\eqref{eq:P1}. Each branch of the conditional corresponds to a partial gradient step along one coordinate, with the margin update playing the role of a slack variable when further optical stabilization is infeasible. In the high-margin regime, the objective is log-convex in each variable separately, and the algorithm converges to a stationary point of \eqref{eq:P1}.}

\subsection{Bi-Directional Outage Probability}
\label{subsec:bidirectional_outage}

For the forward link (A $\to$ B), the stability parameters are
$\phi_{\text{tx},A} = \theta_{\text{div},A}^2/(4\sigma_A^2)$ and
$\phi_{\text{rx},B} = \theta_{\text{fov},B}^2/(4\sigma_B^2)$.
Conversely, for the return link (B $\to$ A), the roles are reversed, yielding
$\phi_{\text{tx},B} = \theta_{\text{div},B}^2/(4\sigma_B^2)$ and
$\phi_{\text{rx},A} = \theta_{\text{fov},A}^2/(4\sigma_A^2)$.

\rev{Let $F$ and $R$ denote the outage events on the forward and return links, respectively, with marginal probabilities $P_\text{F}=P(F)$ and $P_\text{R}=P(R)$ given by \eqref{outage_simp} applied separately to each direction. The bi-directional outage event is the union $F\cup R$, with
\begin{equation}
	P_\text{out,bi}=P(F\cup R)=P_\text{F}+P_\text{R}-P(F\cap R),
\end{equation}
and is therefore determined by the joint distribution of $F$ and $R$, not by the marginals alone.}

\rev{The pointing-error components associated with the two directions are not necessarily independent. Terminal~A's residual angular misalignment contributes to both the forward-link transmitter-side loss and the return-link receiver-side loss, and the same coupling applies to Terminal~B. This is because the transmit and receive functions may share common spacecraft attitude errors, optical-bench disturbances, and PAT-loop dynamics. A large attitude excursion at one terminal can therefore degrade both directions simultaneously, so $F$ and $R$ are physically expected to be positively correlated rather than independent. Capturing the exact dependence would require a specific cross-coupling model between transmit and receive functions at each terminal. Instead, we obtain a model-free range by bracketing $P_\text{out,bi}$ using the Fr\'echet--Hoeffding bounds \cite{frechet1935generalisation}.}

\rev{For two events with given marginal probabilities $P_\text{F}$ and $P_\text{R}$, the Fr\'echet--Hoeffding bounds imply
\begin{equation}
	\label{eq:frechet_full}
	\max(P_\text{F},P_\text{R})
	\le
	P_\text{out,bi}
	\le
	\min(1,P_\text{F}+P_\text{R}).
\end{equation}
The lower bound is attained when the two events are nested, corresponding to the strongest positive dependence allowed by the two marginal probabilities. The upper bound is attained under the most adverse negative-dependence structure allowed by the two marginals, where the overlap $P(F\cap R)$ is minimized.}

\rev{For the OISL setting considered here, such extreme negative dependence is not physically representative, since common attitude excursions, structural vibrations, and PAT-loop disturbances tend to degrade both link directions rather than improving one direction while degrading the other. We therefore restrict attention to nonnegatively correlated outage events, for which
\begin{equation}
	P(F\cap R)\ge P_\text{F}P_\text{R}.
\end{equation}
Under this physically motivated restriction, the upper bound in \eqref{eq:frechet_full} tightens to the independence value $P_\text{F}+P_\text{R}-P_\text{F}P_\text{R}$. The physically admissible envelope is therefore
\begin{equation}
	\label{eq:frechet_hoeffding}
	\max(P_\text{F},P_\text{R})
	\le
	P_\text{out,bi}
	\le
	P_\text{F}+P_\text{R}-P_\text{F}P_\text{R}.
\end{equation}
The lower bound corresponds to fully nested outage events, i.e., a common disturbance triggers both directions as strongly as allowed by the marginal probabilities, whereas the upper bound corresponds to statistically independent outage events. Equation~\eqref{eq:frechet_hoeffding} is sharp under the restriction $P(F\cap R)\ge P_\text{F}P_\text{R}$; every joint distribution with marginals $P_\text{F}$ and $P_\text{R}$ satisfying this restriction produces a $P_\text{out,bi}$ inside this interval.}

\rev{The width of the envelope can be quantified by examining the ratio between the independence upper bound and the fully correlated lower bound. For the symmetric case $P_\text{F}=P_\text{R}=P$, the bounds reduce to $P$ and $2P-P^2$, respectively, so
\begin{equation}
	\label{eq:fh_ratio}
	\frac{P_\text{out,bi}^{\text{(ind)}}}{P_\text{out,bi}^{\text{(corr)}}}
	=
	2-P
	\longrightarrow 2,
	\quad
	P\to 0.
\end{equation}
For the asymmetric case $P_\text{F}\ne P_\text{R}$, assuming without loss of generality that $P_\text{F}\ge P_\text{R}$, the ratio becomes
\begin{equation}
	\frac{P_\text{F}+P_\text{R}-P_\text{F}P_\text{R}}{P_\text{F}}
	=
	1+\frac{P_\text{R}}{P_\text{F}}(1-P_\text{F}),
\end{equation}
which tends to $1+P_\text{R}/P_\text{F}$ as both outage probabilities approach zero. This ratio is upper bounded by $2$ and reaches this upper bound only in the symmetric limit. Hence, in the high-margin regime where $P_\text{F},P_\text{R}\ll1$, the independence assumption overestimates the bi-directional outage by at most a factor of two relative to the fully correlated lower bound, with the gap decreasing as the two link directions become more asymmetric.}

\rev{This factor-of-two outage gap can be translated into a link-margin penalty using the asymptotic decay law $P_\text{out}\propto M^{-\phi_\text{min}}$, where $M$ denotes the linear link margin. Designing for the independence upper bound rather than the fully correlated lower bound requires a margin increase of
\begin{equation}
	\label{eq:fh_margin_penalty}
	\Delta M_\text{worst}
	=
	\frac{10\log_{10}2}{\phi_\text{min}}~\text{dB}
	\approx
	\frac{3.01}{\phi_\text{min}}~\text{dB}.
\end{equation}
This penalty is at most approximately $3$~dB at the marginal-stability boundary $\phi_\text{min}=1$ and decreases as $1/\phi_\text{min}$ for higher-stability terminals. For example, it is below $1$~dB for $\phi_\text{min}\ge3$ and below $0.3$~dB for the reference design of Table~\ref{tab:linkbudget}. The Fr\'echet--Hoeffding envelope is therefore particularly tight in the high-stability operating regime targeted in this paper, justifying the use of the strict-independence expression as a sharp upper bound for design purposes under nonnegative outage dependence.}

\rev{Although the prefactor of $P_\text{out,bi}$ can vary by up to a factor of two across the admissible envelope, the asymptotic decay exponent is invariant. Both the fully correlated lower bound and the independence upper bound scale as $M^{-\phi_\text{min,bi}}$, where
\begin{equation}
	\phi_\text{min,bi}
	=
	\min\left(
	\phi_{\text{tx},A},
	\phi_{\text{rx},B},
	\phi_{\text{tx},B},
	\phi_{\text{rx},A}
	\right)
\end{equation}
is the global weakest-link stability parameter. The weakest-link principle of Section~\ref{sec:asymptotic} therefore extends to the bi-directional setting without modification, regardless of the precise nonnegative dependence structure. A refined analysis based on a common-mode/differential-mode decomposition of the terminal-attitude jitters is identified as future work.}

\rev{For the subsequent design discussion, we adopt the upper-bound expression in \eqref{eq:frechet_hoeffding}, which corresponds to the strict-independence case and represents a sharp upper bound on the true $P_\text{out,bi}$ under any nonnegatively correlated dependence structure:}
\begin{equation}
	P_{\text{out,bi}}
	=
	1-\bigl(1-P_{\text{out},A\to B}\bigr)
	\bigl(1-P_{\text{out},B\to A}\bigr).
\end{equation}
In the high-margin regime, where $P_{\text{out}}\ll1$, this is well approximated by the union expression
\begin{equation}
	P_{\text{out,bi}}
	\approx
	P_{\text{out},A\to B}
	+
	P_{\text{out},B\to A}.
\end{equation}
Because the sum is asymptotically dominated by the term with the slowest decay rate, the overall system decay exponent $d_{\text{bi}}$ is dictated by the global weakest link:
\begin{equation}
	d_{\text{bi}}
	=
	\min\bigl(
	\phi_{\text{tx},A},
	\phi_{\text{rx},B},
	\phi_{\text{tx},B},
	\phi_{\text{rx},A}
	\bigr).
\end{equation}

\subsection{Achievable Symmetric Data Rate}
The achievable symmetric data rate is also constrained by the direction with lower ergodic capacity. \rev{Using the high-SNR equivalent-SNR mapping obtained by matching $\log_2(\gamma_\text{eff})=\log_2(\gamma)+\Delta C_\text{jitter}$ with $\Delta C_\text{jitter} = -(\xi/\ln 2)S$ from \eqref{eq:capacity_penalty}, where $S=1/\phi_\text{tx}+1/\phi_\text{rx}$, conversion from $\log_2$ to $\ln$ gives $\ln(\gamma_\text{eff})=\ln(\gamma) - \xi S$. The effective high-SNR equivalent SNR in each direction is therefore}
\begin{equation}
	\label{eq:gamma_eff}
	\gamma_{\text{eff}, A\to B} \approx \gamma \cdot \exp\!\left( - \xi\!\left( \frac{1}{\phi_{\text{tx},A}} + \frac{1}{\phi_{\text{rx},B}} \right) \right),
\end{equation}
and similarly for $B\to A$. \footnote{\rev{We emphasize that $\gamma_\text{eff}$ is a high-SNR equivalent-SNR construct (slope-matched to the asymptotic ergodic-capacity penalty), not an exact finite-SNR (at moderate SNR the exact integral \eqref{eq:exact_ergodic_capacity} should be used).}} The achievable symmetric data rate $R_{\text{sym}}$ is therefore limited by the weaker direction:
\begin{equation}
	R_{\text{sym}} = \min\bigl( C_{A\to B}, C_{B\to A} \bigr),
\end{equation}
where $C_{A\to B} = \log_2(1 + \gamma_{\text{eff}, A\to B})$.

\subsection{Design Optimization and Tradeoffs for Asymmetric Terminals}
To maximize both reliability and symmetric throughput in heterogeneous links, such as those between a highly stable geostationary (GEO) relay with small $\sigma_{B}$ and a jitter‑prone low Earth orbit (LEO) small satellite with large $\sigma_{A}$, the optical parameters must be allocated so that the stability is balanced across all four functions of the bidirectional link. Setting $\phi_{\text{tx},A} = \phi_{\text{tx},B}$ yields the optimum beamwidth allocation rule
\begin{equation}
	\label{eq:theta_ratio}
	\frac{\theta_{\text{div},A}}{\theta_{\text{div},B}} = \frac{\sigma_A}{\sigma_B}.
\end{equation}
Similarly, the receiver FOV must be scaled as $\theta_{\text{fov},A} / \theta_{\text{fov},B} = \sigma_A / \sigma_B$. \rev{Equation~\eqref{eq:theta_ratio} is the simultaneous-stationarity condition of Problem~(P1) (cf.\ Section~\ref{subsec:opt_problem}) applied at both directions, rather than a separately asserted optimum.}

However, to operationalize the stability-balancing rule in \eqref{eq:theta_ratio}, it is necessary to quantify the physical and power costs incurred by beam widening at the less stable terminal. \rev{Since the on-axis transmitter gain satisfies $G_\text{tx,peak}\propto\theta_\text{div}^{-2}$, widening the beam at the less stable terminal reduces its on-axis gain and therefore requires a proportionally larger transmit power to maintain the same received power under perfect alignment.} From the truncated‑Gaussian model, the peak gain scales as $G_{\text{tx}} \propto \theta_{\text{div}}^{-2}$ \rev{via the optimally-truncated Klein--Degnan relation $D_\text{tx}\propto 1/\theta_\text{div}$ combined with $G_\text{tx0}=\pi D_\text{tx}^2/\lambda^2$} \cite{Antena-gain}. Consequently, if the beamwidth of the LEO satellite is increased by a factor $\kappa = \sigma_A/\sigma_B$ relative to that of the GEO satellite, its transmit power must be increased by the same factor $\kappa^2$ to maintain the same received power under perfect alignment. More generally, for a link with asymmetric stabilities, the required transmit power ratio between the two terminals is
\begin{equation}
	\label{eq:Ptx_ratio}
	\frac{P_{\text{tx},A}}{P_{\text{tx},B}} = \left( \rev{\frac{\theta_{\text{div},A}}{\theta_{\text{div},B}}} \right)^2 = \left( \rev{\frac{\sigma_A}{\sigma_B}} \right)^2.
\end{equation}
This relation exposes a fundamental energy–stability trade‑off. That is, a satellite with poor pointing stability must either expend significantly more transmit power or accept a lower link margin (hence higher outage) than a stable counterpart. The optimization framework therefore provides a quantitative basis for trading off power amplifier size, solar panel capacity, and battery mass against pointing control precision in the early stages of satellite design.

\section{Numerical Results and Design Guidelines}
\label{sec:numerical}

In this section, we validate the analytical expressions derived in Sections~\ref{III-stat}, \ref{sec:outage}, and \ref{sec:asymptotic} using both analytical evaluation and Monte Carlo simulations, and use the resulting curves to extract practical design guidelines for OISL parameter selection. Unless stated otherwise, the simulations assume an operating wavelength of $\lambda = 1550~\text{nm}$ and a link distance of $z = 1000~\text{km}$, independent Rayleigh pointing jitter at the transmitter and receiver with standard deviations $\sigma_{\text{tx}}$ and $\sigma_{\text{rx}}$, and Gaussian approximations for both the transmit far field pattern and the receiver coupling efficiency.

\rev{Empirical estimates of $P_\text{out}$ in Figs.~\ref{fig:outage_slope}, \ref{fig:outage_slope_gain}, and \ref{fig:beamwidth_optimization} are obtained from $5\times 10^{7}$ independent channel realizations per operating point; at outage levels below $10^{-4}$. Analytical curves below the empirical floor of $P_\text{out}\approx 10^{-5}$ are drawn in a lighter line style and labeled as ``closed-form extrapolation,'' to distinguish simulation-supported from analytically extrapolated regions. Analytical curves are computed using \eqref{main-pdf}, \eqref{outage_simp}, and \eqref{outage_symetric}.} Additional simulation parameters are stated in the caption of each figure. The beam divergence and receiver FOV are varied to span practically relevant stability regimes, and results are reported in terms of the dimensionless stability parameters $\phi_{\text{tx}}$ and $\phi_{\text{rx}}$ to preserve generality across different optical hardware configurations.

\begin{figure*}
	\centering
	\subfloat[] {\includegraphics[width=2.32 in]{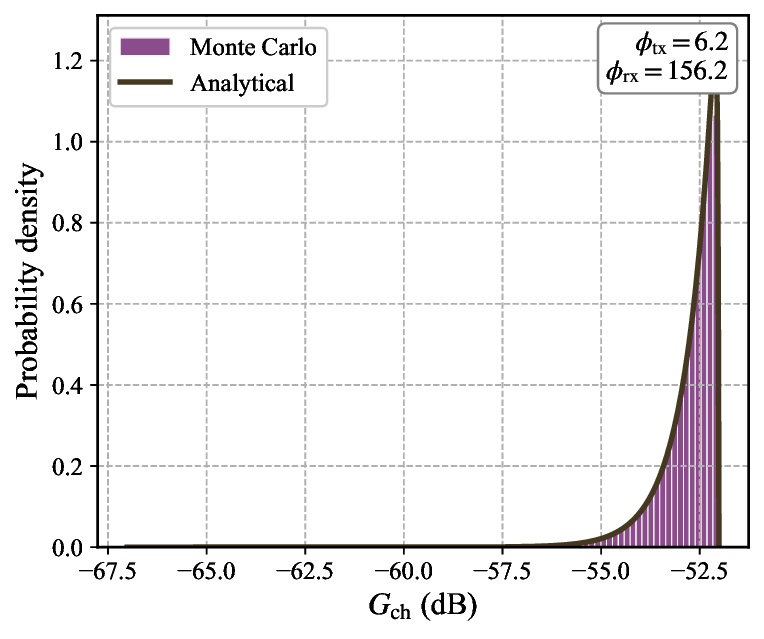}
		\label{cf1}
	}
	\hfill
	\subfloat[] {\includegraphics[width=2.32 in]{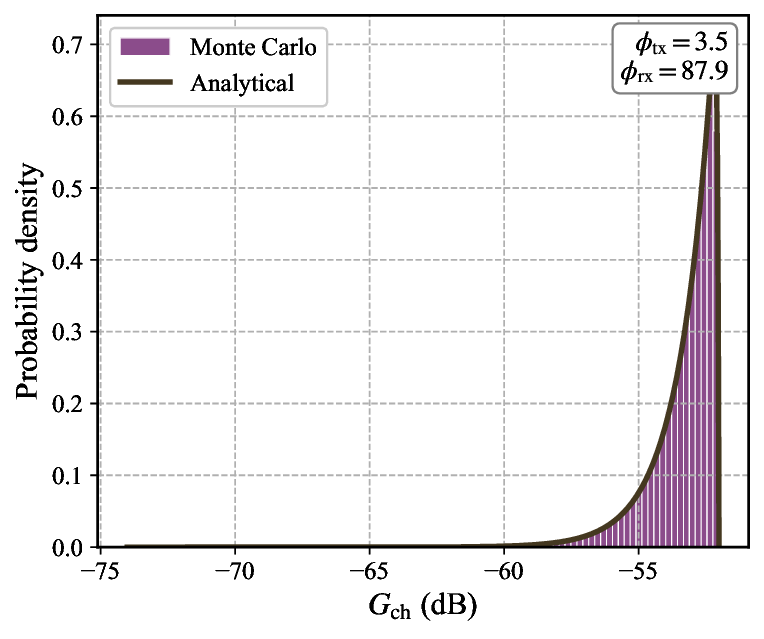}
		\label{cf2}
	}
	\hfill
	\subfloat[] {\includegraphics[width=2.32 in]{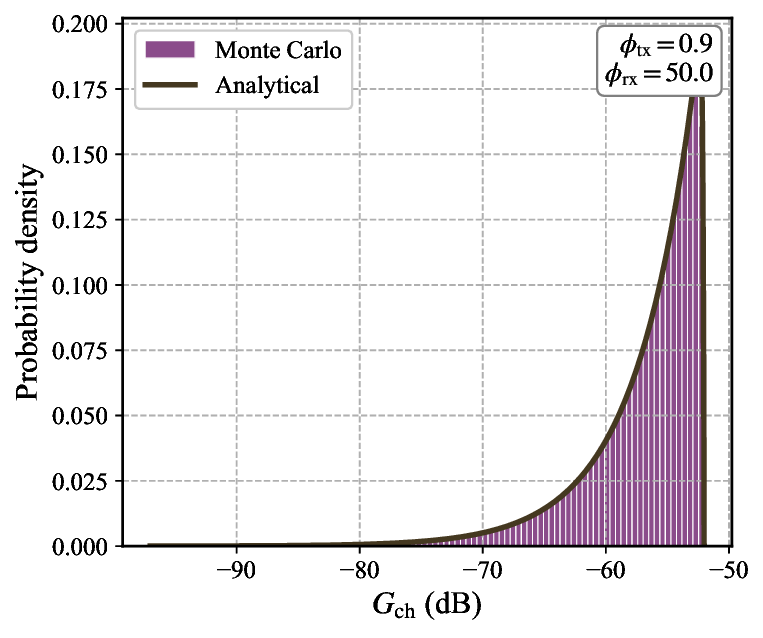}
		\label{cf3}
	}

	\caption{Comparison between the analytical PDF of the instantaneous channel gain and Monte Carlo simulations \rev{of the Gaussian--Rayleigh channel model} for three representative stability configurations: (a) highly imbalanced but stable link, (b) moderately imbalanced link, and (c) unstable regime with $\phi_{\text{tx}}<1$. \rev{The close agreement validates the closed-form PDF expressions \eqref{main-pdf} and \eqref{special-pdf} as representations of the underlying Gaussian--Rayleigh model from which they are derived; the quantitative fidelity of the Gaussian--Rayleigh model itself relative to the exact diffraction-optics response is calibrated separately in Section~\ref{subsubsec:exact_mc} (Table~\ref{tab:exact_mc}).} \rev{The horizontal axis is in dB, $X=10\log_{10}(G_\text{ch})$; the analytical curves are accordingly transformed via $f_X(x) = f_{G_\text{ch}}(10^{x/10})(\ln 10/10)\,10^{x/10}$ so that both curves and histograms are densities of the same dB-valued random variable.} Other simulation parameters: $D_\textrm{tx} = D_\textrm{rx}$ = 10 cm, $\eta_\text{tx} = 0.7, \eta_\text{rx} = 0.7, \eta_{\text{opt}} = 0.5$.}

	\label{pdf-comparison}
\end{figure*}

\rev{The validation strategy adopted in this section proceeds at two complementary levels. First, in Section~\ref{subsec:validation_analytical} (Fig.~\ref{pdf-comparison}) and in the parametric studies of Sections~\ref{subsec:impact_stability}--\ref{subsec:Bi-Directional Asymmetric Link} (Figs.~\ref{fig:outage_slope}--\ref{fig:beamwidth_optimization}), we validate the closed-form analytical expressions derived in Sections~\ref{III-stat}--\ref{sec:asymptotic} against Monte Carlo simulations of the Gaussian--Rayleigh channel model, confirming the correctness of the analytical derivations. Second, in Section~\ref{subsubsec:exact_mc} (Table~\ref{tab:exact_mc}), we validate the Gaussian--Rayleigh channel model itself against the exact diffraction-optics response (Klein--Degnan transmitter and Airy-on-detector receiver), establishing the regime of quantitative tightness of the overall framework. The qualitative design principles extracted in subsequent steps, i.e., the weakest-link principle, the role of the asymptotic power offset, the beamwidth-stability tradeoff, and the capacity-penalty saturation, are intrinsic features of the Gaussian--Rayleigh model that remain qualitatively faithful to the exact diffraction response across the full range of $\phi$. Absolute margin values at low-stability operating points should be calibrated via the offsets reported in Table~\ref{tab:exact_mc}.}

\rev{\subsection{Reference Link Budget and Threshold Mapping}
\label{subsec:linkbudget}
To make the abstract threshold $G_\text{th}$ concretely interpretable, we report in Table~\ref{tab:linkbudget} a complete reference link budget for an IM/DD APD--OOK OISL operating at $10$~Gbps. The receiver employs an InGaAs avalanche photodiode (APD) with responsivity $\mathcal{R}=0.9$~A/W, multiplication factor $M_\text{apd}=10$, excess noise factor $F(M_\text{apd})=4.5$, and a transimpedance amplifier with input-referred current noise density $i_n=20$~pA$/\sqrt{\text{Hz}}$. Background light is assumed negligible owing to the shielded space-vacuum environment. Under the Gaussian approximation for OOK detection, and defining the electrical SNR as
\begin{equation}
    \gamma=\frac{(I_1-I_0)^2}{\sigma_1^2+\sigma_0^2},
\end{equation}
the SNR threshold for a target pre FEC BER of $10^{-3}$ is $\gamma_\text{th}=2Q_0^2$, where $Q_0=Q^{-1}(10^{-3})\approx 3.09$. This gives $\gamma_\text{th}\approx 19.1$, or equivalently $12.8$~dB. The corresponding receiver sensitivity is $P_\text{r,th} \approx -31.6$~dBm. Therefore, the threshold channel gain that the link must deliver is
\begin{equation}
	\label{eq:Gth_mapping}
	G_\text{th} = \frac{P_\text{r,th}}{P_\text{tx,max}}\bigg|_\text{linear},
\end{equation}
which evaluates to $G_\text{th}\approx -61.6$~dB for $P_\text{tx,max}=30$~dBm. The corresponding peak channel gain is $G_\text{peak}\approx -53.65$~dB after accounting for $\eta_T\approx -0.89$~dB transmitter taper loss, $\eta_R\approx -0.76$~dB on-axis Airy-to-detector spillover, lumped optical efficiency $\eta_\text{opt}=0.5$, and surface efficiencies $\eta_\text{tx}=\eta_\text{rx}=0.7$. The resulting reference link margin is $M_\text{dB}=G_\text{peak,dB}-G_\text{th,dB}\approx +7.95$~dB. The reference operating point of Table~\ref{tab:linkbudget} yields stability parameters $\phi_\text{tx}\approx 13.3$ and $\phi_\text{rx}\approx 39.1$, which together with this link margin produce an outage probability of approximately $1.87\times 10^{-10}$ from \eqref{outage_simp}. This indicates that the link is highly unlikely to fall below the channel gain required to satisfy the adopted $10^{-3}$ pre FEC BER target, consistent with the high-availability operating regime targeted by modern commercial OISL terminals.}


\begin{table}[!t]
\centering
\rev{
\caption{\rev{Reference link budget for the IM/DD APD--OOK OISL used in Section~\ref{sec:numerical}.}}
\label{tab:linkbudget}
\footnotesize
\renewcommand{\arraystretch}{1.15}
\setlength{\tabcolsep}{4pt}
\begin{tabular}{lll}
\toprule
\textbf{Parameter} & \textbf{Symbol} & \textbf{Value} \\
\midrule
Wavelength               & $\lambda$            & $1550$~nm \\
Link distance            & $z$                  & $1000$~km \\
TX/RX aperture diameter  & $D_\text{tx}, D_\text{rx}$ & $10$~cm \\
Obscuration / truncation & $\gamma_o, \alpha_0$ & $0$~/~$1.12$ \\
Beam divergence          & $\theta_\text{div}$  & $14.6\,\mu$rad \\
Equivalent Gaussian FOV  & $\theta_\text{fov}$  & $25\,\mu$rad \\
TX/RX pointing jitter    & $\sigma_\text{tx},\sigma_\text{rx}$ & $2\,\mu$rad \\
Data rate & $R_b$ & $10$~Gbps \\
Electrical bandwidth & $B_\text{el}$ & $R_b/2=5$~GHz \ \\
Transmit power           & $P_\text{tx,max}$    & $30$~dBm  \\
\midrule
Modulation               & --                   & OOK \\
Detector                 & --                   & InGaAs APD direct det. \\
APD responsivity         & $\mathcal{R}$        & $0.9$~A/W \\
APD gain / excess noise factor  & $M_\text{apd}, F$    & $10$, $4.5$ \\
TIA input noise density  & $i_n$                & $20$~pA$/\sqrt{\text{Hz}}$ \\
Thermal noise (RMS)      & $\sigma_\text{th}$   & $1.4\,\mu$A \\
Pre-FEC BER target       & BER                  & $10^{-3}$ \\
OOK Q-factor threshold                 & $Q_0$                & $3.09$ \\
\midrule
TX taper efficiency      & $\eta_T$             & $-0.89$~dB \\
RX on-axis spillover     & $\eta_R$             & $-0.76$~dB \\
Lumped efficiencies      & $\eta_\text{tx},\eta_\text{rx},\eta_\text{opt}$ & $0.7,0.7,0.5$ \\
\midrule
SNR threshold            & $\gamma_\text{th}$   & $+12.8$~dB \\
Receiver sensitivity     & $P_\text{r,th}$      & $-31.6$~dBm \\
Threshold channel gain   & $G_\text{th}$        & $-61.6$~dB \\
Peak channel gain        & $G_\text{peak}$      & $-53.65$~dB \\
Link margin              & $M_\textrm{dB}$                  & $+7.95$~dB \\
\midrule
Stability parameters     & $\phi_\text{tx}, \phi_\text{rx}$ & $13.3, 39.1$ \\
Outage at reference op.\ pt.\ & $P_\text{out}$ & $\approx 1.87\times 10^{-10}$ \\
\bottomrule
\end{tabular}
}
\end{table}

\subsection{Validation of Analytical Models}
\label{subsec:validation_analytical}
Fig.~\ref{pdf-comparison} compares the analytical PDF derived in Section~\ref{III-stat} with Monte Carlo simulations \rev{of the Gaussian--Rayleigh channel model} for three representative pairs of stability parameters. Several observations can be made from this figure. First, in all cases, the analytical curves closely match the simulated histograms across the entire dynamic range. \rev{The close agreement confirms that the closed-form PDF expressions \eqref{main-pdf} and \eqref{special-pdf} faithfully represent the Gaussian--Rayleigh channel model from which they are derived, including in the unstable regime $\phi_\text{tx}<1$ (case (c)). This validates the correctness of the derivations underlying the outage probability analysis and the asymptotic design insights of the paper. The quantitative fidelity of the Gaussian--Rayleigh model itself relative to the exact diffraction-optics response is a distinct question, addressed separately in Section~\ref{subsubsec:exact_mc} via direct Monte Carlo propagation through the Klein--Degnan and Airy-on-detector responses.}

Second, each case shows different channel characteristics. Particularly, case~(a) corresponds to a highly imbalanced but overall stable link, where the receiver is significantly more stable than the transmitter ($\phi_{\text{tx}} = 6.2$, $\phi_{\text{rx}} = 156.2$). This configuration yields a PDF that is sharply concentrated near the peak gain and exhibits a steep roll-off toward deep fades. In case~(b), the imbalance is reduced but the transmitter remains the less stable terminal ($\phi_{\text{tx}} = 3.5$, $\phi_{\text{rx}} = 87.9$). The resulting distribution has a heavier tail toward lower gains and an increased probability of moderate fades. Case~(c) illustrates an unstable regime with $\phi_{\text{tx}} < 1$, where the PDF displays a pronounced heavy tail toward deep fades, indicating a high likelihood of severe pointing-induced attenuation even when the opposite terminal remains relatively stable.

\subsubsection{Approximation Accuracy and Operational Boundaries}
\label{subsubsec:approx_accuracy}
While Figs.~\ref{fig:ffip_validation_tx_side} and \ref{fig:ffip_validation_rx_side} show close agreement within the main lobe, the accuracy of the outage prediction \rev{of the Gaussian--Rayleigh model relative to the exact diffraction-optics response} depends on the probability that the pointing error exceeds the validity range of the Gaussian approximation. \rev{We define the operational validity boundary at each terminal as $\theta_\text{valid,i} = \beta_i\,\theta_{\text{ref},i}$, where $\theta_{\text{ref},\text{tx}}=\theta_\text{div}$, $\theta_{\text{ref},\text{rx}}=\theta_\text{fov}$, and $\beta_i\in(0,1]$ is the normalized validity radius read from Figs.~\ref{fig:ffip_validation_tx_side}--\ref{fig:gauss_error_db}. The demonstrated radii are $\beta_\text{tx}\approx 0.7$ (sub-$1$~dB Gaussian-vs-exact transmitter error across $\gamma_o\in[0,0.2]$, with the worst case $\gamma_o=0$ reaching $\approx 0.8$~dB and centrally obscured configurations giving strictly smaller errors) and $\beta_\text{rx}\approx 0.3$ (sub-$0.5$~dB Gaussian-vs-exact receiver coupling error across all tested detector geometries).}

To quantify the impact of this deviation on the outage probability $P_{\text{out}}$, we evaluate the probability that the instantaneous pointing error $\Theta$ exceeds the validity region \rev{($\Theta > \beta_i\,\theta_{\text{ref},i}$)}. Using the Rayleigh cumulative distribution implied by \eqref{pointing-pdf1}, this probability is
\begin{equation}
	\label{eq:prob_invalid}
	\rev{P(\Theta_i > \beta_i\theta_{\text{ref},i}) = \exp\!\left( -\frac{\beta_i^2\,\theta_{\text{ref},i}^{2}}{2 \sigma_{i}^{2}} \right).}
\end{equation}
Substituting the stability parameter from \eqref{stability_parameters} yields
\begin{equation}
	\rev{P_{\text{invalid},i} = e^{-2\beta_i^2\phi_i}.}
\end{equation}
\rev{In particular, on the transmitter side $P_{\text{invalid},\text{tx}} = e^{-0.98\,\phi_\text{tx}}$ and on the receiver side $P_{\text{invalid},\text{rx}} = e^{-0.18\,\phi_\text{rx}}$.} This result shows that the approximation error is exponentially suppressed as link stability increases. \rev{Considering the desired outage probability, two regimes are of practical interest.}

\begin{itemize}
	\item {\rev{High-stability regime ($\phi_\text{tx}\ge 7$ and $\phi_\text{rx}\ge 38$):}}
	\rev{In this regime, the probability of leaving the Gaussian validity region is bounded by $P_{\text{invalid},\text{tx}}\le e^{-6.86}\approx 1.05\times 10^{-3}$ and $P_{\text{invalid},\text{rx}}\le e^{-6.84}\approx 1.07\times 10^{-3}$ on the two sides. The deep fades that determine outage at $P_\text{out}\sim 10^{-3}$ occur almost entirely within the main lobe, where the Gaussian approximation remains highly accurate (error $<1$~dB on the transmitter side and $<0.5$~dB on the receiver side). The Reference design of Table~\ref{tab:linkbudget} satisfies this regime by a comfortable margin ($\phi_\text{tx}=13.3$, $\phi_\text{rx}=39.1$), and the resulting closed-form prediction is tight relative to the exact diffraction response, as confirmed by Table~\ref{tab:exact_mc} below.}

	\item {Lower-stability regime \rev{($\phi<7$ on the TX or $\phi<38$ on the RX)}:}
	\rev{Below the high-stability bound, the probability of leaving the Gaussian validity region is no longer negligible, and the closed-form prediction departs from the exact-model outage by several decibels of required margin. The qualitative dependence of outage on the stability parameters (polynomial slope governed by $\phi_\text{min}$, power offset shaped by $\phi_\text{max}$) is preserved, but absolute margins should be calibrated using the offsets in Table~\ref{tab:exact_mc}. As a representative case, at $\phi=0.5$ the invalidity probability rises to $e^{-0.49}\approx 61\%$ on the transmitter side and $e^{-0.09}\approx 91\%$ on the receiver side, and the sign of the Gaussian-vs-exact margin discrepancy becomes regime-dependent. Particularly, in the moderate-margin regime, the Airy and Bessel main-lobe coupling lies below the curvature-matched Gaussian, so the closed-form under-predicts the required margin (the Gaussian is optimistic). In contrast, at extreme margins, the diffraction sidelobes of the exact response provide a coupling floor that the Gaussian tail does not capture, restoring conservatism. The exact-model Monte Carlo validation of Section~\ref{subsubsec:exact_mc} quantifies these deviations across representative configurations.}
\end{itemize}

\rev{The high-stability bound above is obtained by requiring $P_{\text{invalid},i}\le 10^{-3}$ on both sides, i.e., $\phi_\text{rx} > \ln(10^3)/(2\beta_\text{rx}^2) \approx 38$ on the receiver side and $\phi_\text{tx}>\ln(10^3)/(2\beta_\text{tx}^2)\approx 7$ on the transmitter side; the receiver-side bound is the binding constraint due to the smaller $\beta_\text{rx}$. The Reference design of Table~\ref{tab:linkbudget} satisfies both bounds.}

\rev{Even when $\phi$ falls below unity, the analytical outage expression \eqref{outage_simp} continues to faithfully represent the Gaussian--Rayleigh channel model from which it is derived: as shown in Fig.~\ref{pdf-comparison}(c) for $\phi_{\text{tx}} = 0.9$, the analytical PDF still matches the Gaussian-MC histogram closely.} This derivation-level robustness arises from two factors:
(i) outage probability is an integrated metric, so moderate inaccuracies in the extreme tail contribute little to the cumulative probability for practical thresholds $G_{\text{th}}$;
(ii) for finite link margins ($M = G_{\text{peak}} / G_{\text{th}} < 30$~dB), the threshold $G_{\text{th}}$ lies above the region where approximation errors are most pronounced. 

\rev{\subsubsection{Exact-Model Monte Carlo Validation}
\label{subsubsec:exact_mc}
To validate the closed-form Gaussian prediction directly against the exact diffraction-optics response, we compare the analytical outage \eqref{outage_simp} against Monte Carlo simulations in which Rayleigh-distributed pointing errors are propagated through the exact Klein--Degnan transmitter response \eqref{eq:ffip} and the exact two-dimensional Airy-on-detector coupling \eqref{eq:coupling}. Pre-computed lookup tables of $\widetilde{L}_\text{tx}^\text{exact}(\theta/\theta_\text{div})$ (from \eqref{eq:ffip} with $\alpha_0=1.12,\gamma_o=0$) and $\widetilde{L}_\text{rx}^\text{exact}(\theta/\theta_\text{fov})$ (from \eqref{eq:coupling} with $r_d/r_\text{Airy}=1$, curvature-matched to $\theta_\text{fov}$ at the origin) reduce per-sample evaluation to interpolation, enabling $N=2\times 10^7$ samples per configuration.  Table~\ref{tab:exact_mc} reports the resulting dB margin error $\Delta M=10\log_{10}(M_\text{Gauss}/M_\text{exact})$ at $P_\text{out}\in\{10^{-3},10^{-5}\}$ for four representative stability configurations chosen to span the validity range of Section~\ref{subsubsec:approx_accuracy}. Positive $\Delta M$ indicates that the closed-form prediction is conservative (over-estimates the required margin); negative $\Delta M$ indicates it is optimistic (under-estimates the required margin).

For the Reference design of Table~\ref{tab:linkbudget}, which satisfies the high-stability bound, the closed-form prediction agrees with the exact-model simulation to within $0.3$~dB at both outage levels and is mildly conservative ($\Delta M=+0.2$~dB at $10^{-3}$, $+0.3$~dB at $10^{-5}$). For configurations that violate the bound (rows 2--4), the Gaussian approximation departs more substantially. In particular, at the moderate-balanced operating point ($\phi_\text{tx}=\phi_\text{rx}=4$), the closed-form under-predicts the required margin by approximately $7$~dB at $P_\text{out}=10^{-3}$, and the deviation grows further as the weaker stability parameter falls below unity. This behavior originates from the residual curvature-match mismatch of the equivalent-Gaussian model. More precisely, within the main-lobe region $|u|\lesssim 0.3\,\theta_\text{ref}$ the Gaussian closely tracks the exact response, but at larger $u$ the Airy main-lobe coupling falls off significantly faster than the curvature-matched Gaussian (Fig.~\ref{fig:gauss_error_db}), causing more probability mass to lie below the outage threshold in the exact channel than the closed-form predicts. The Degraded-RX entry at $P_\text{out}=10^{-5}$ ($\Delta M=+1.0$~dB) shows the opposite-sign deviation that arises in the extreme-margin regime where the exact-response diffraction sidelobes provide a coupling floor that the Gaussian tail does not capture. This is a curiosity of the deep-tail regime rather than a design-relevant effect. Overall, Table~\ref{tab:exact_mc} confirms that the closed-form Gaussian framework is accurate and mildly conservative within the high-stability bound $\phi_\text{rx}>38$ targeted by modern high-reliability OISL terminals, while remaining qualitatively faithful (i.e., correctly capturing weakest-link slope, power-offset behavior, and the existence of a beamwidth optimum) outside this bound.

\begin{table}[!t]
\centering
\caption{\rev{Margin error $\Delta M=10\log_{10}(M_\text{Gauss}/M_\text{exact})$ of the closed-form Gaussian prediction versus an exact Klein--Degnan/Airy Monte Carlo simulation, at two outage levels and four stability configurations. Positive $\Delta M$ indicates conservative (over-estimating the required margin), negative indicates optimistic. The first row corresponds to the Reference design of Table~\ref{tab:linkbudget}, which satisfies the high-stability bound $\phi_\text{rx}>38$ of Section~\ref{subsubsec:approx_accuracy}; the remaining rows progressively violate this bound.}}
\label{tab:exact_mc}
\footnotesize
\renewcommand{\arraystretch}{1.2}
\setlength{\tabcolsep}{6pt}
\begin{tabular}{lcccc}
\toprule
\textbf{Configuration} & $\phi_\text{tx}$ & $\phi_\text{rx}$ & $\Delta M\,[10^{-3}]$ & $\Delta M\,[10^{-5}]$ \\
\midrule
Reference (Table~\ref{tab:linkbudget}) & $13.3$ & $39.1$ & $+0.2$~dB  & $+0.3$~dB \\
Moderate balanced                       & $4.0$  & $4.0$  & $-6.9$~dB  & $-6.7$~dB \\
Degraded RX                             & $13.3$ & $2.0$  & $-5.3$~dB  & $+1.0$~dB \\
TX-limited imbalanced                   & $1.0$  & $20.0$ & $-18.1$~dB & $-28.5$~dB \\
\bottomrule
\end{tabular}
\end{table}
}

\rev{Finally, based on what we have discussed in this subsection, the two-level validation strategy is completed. Fig.~\ref{pdf-comparison} confirms that the closed-form expressions of Sections~\ref{III-stat}--\ref{sec:asymptotic} accurately reproduce the Gaussian--Rayleigh channel model, and Table~\ref{tab:exact_mc} calibrates the fidelity of the Gaussian--Rayleigh model itself relative to the exact diffraction-optics response. The framework is tight (sub-decibel margin error) within the high-stability bound $\phi_\text{rx}>38$ that defines the operating regime targeted by modern high-reliability OISL terminals, and remains qualitatively faithful outside this bound. 

The parametric studies of the next subsections deliberately span stability values below this bound to render the polynomial outage-slope and weakest-link phenomena visible on a logarithmic scale. The qualitative trends extracted there are intrinsic features of the Gaussian--Rayleigh model that carry over to the exact response, while absolute margin numbers for low-stability operating points should be interpreted via the offsets in Table~\ref{tab:exact_mc}.}
\subsection{Impact of Stability Parameters}
\label{subsec:impact_stability}
\subsubsection{Outage Probability}
\begin{figure}[htbp]
	\centering
	\includegraphics[width=0.85\columnwidth]{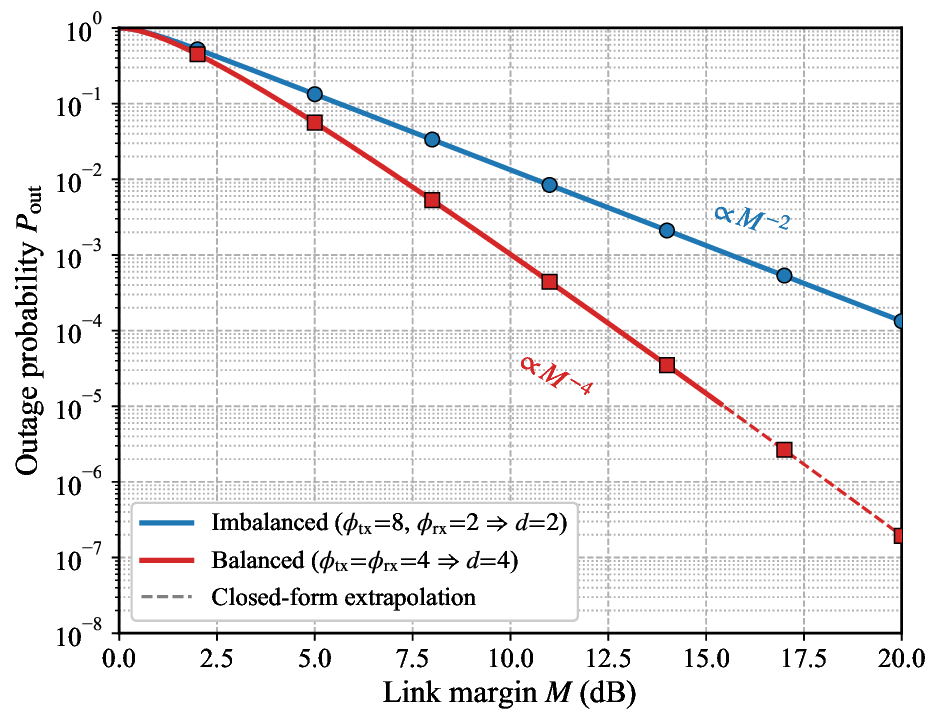}
	\caption{Outage probability versus link margin for balanced and imbalanced stability configurations\rev{, computed from \eqref{outage_simp} and verified against Monte Carlo of the Gaussian--Rayleigh channel model}. The asymptotic slope of each curve is governed by the smaller stability parameter (the weakest link principle). The balanced case achieves a steeper decay despite lower individual peak stability, highlighting the importance of stability parity over unilateral optimization. \rev{Analytical curves below $10^{-5}$ are closed-form extrapolations.}}
	\label{fig:outage_slope}
\end{figure}

Fig.~\ref{fig:outage_slope} illustrates the weakest link principle in OISL design by comparing the outage probability for balanced and imbalanced pointing stability configurations. \rev{The stability values plotted here ($\phi=2,4,8$) are deliberately chosen below the high-stability bound of Section~\ref{subsubsec:approx_accuracy} so that the polynomial-slope behavior $P_\text{out}\propto M^{-\phi_\text{min}}$ is fully visible on the logarithmic scale. Absolute margins for these specific stability values should be read together with the offsets in Table~\ref{tab:exact_mc}.} In the imbalanced case with $\phi_{\text{tx}}=8$ and $\phi_{\text{rx}}=2$, the asymptotic decay of the outage curve is governed entirely by the weaker terminal, resulting in an order of $d=2$. In contrast, the balanced configuration with $\phi_{\text{tx}}=\phi_{\text{rx}}=4$ exhibits a substantially steeper decay with an order of $d=4$, despite each individual terminal having lower peak stability than the stronger terminal in the imbalanced case. This comparison highlights that overall link reliability is fundamentally bottlenecked by the terminal with poorer normalized pointing stability. As a result, for a given link margin, the balanced system achieves orders of magnitude lower outage probability than the imbalanced system, with the performance gap widening rapidly as the margin increases.

These results yield clear design implications. First, when outage probability is the key performance metric, balancing the pointing stability of the transmitter and receiver is more effective than maximizing the stability of only one terminal, as excess stability at one end cannot compensate for the heavy tail induced by the weaker terminal. Second, improvements in outage slope and long term reliability require increasing the stability parameter of the weakest terminal, which can be achieved by reducing platform jitter through improved attitude control and tracking or by appropriately widening the beam divergence or receiver FOV at that terminal. Finally, when the asymptotic decay exponent, $d$, is low, increasing the link margin through transmit power provides a marginal gain, whereas targeted improvements to the pointing stability of the bottlenecked terminal yield substantially more efficient gains in link availability. \rev{All three implications are robust features of the Gaussian--Rayleigh model that carry over qualitatively to the exact diffraction response, since the weakest-link slope is determined by the polynomial tail of the loss distribution \eqref{eq:fLi}, which is itself derived from the universal Rayleigh-to-exponential change of variables and therefore independent of the specific shape of the optical response within its main lobe.}

\begin{figure}[htbp]
	\centering
	\includegraphics[width=0.85\columnwidth]{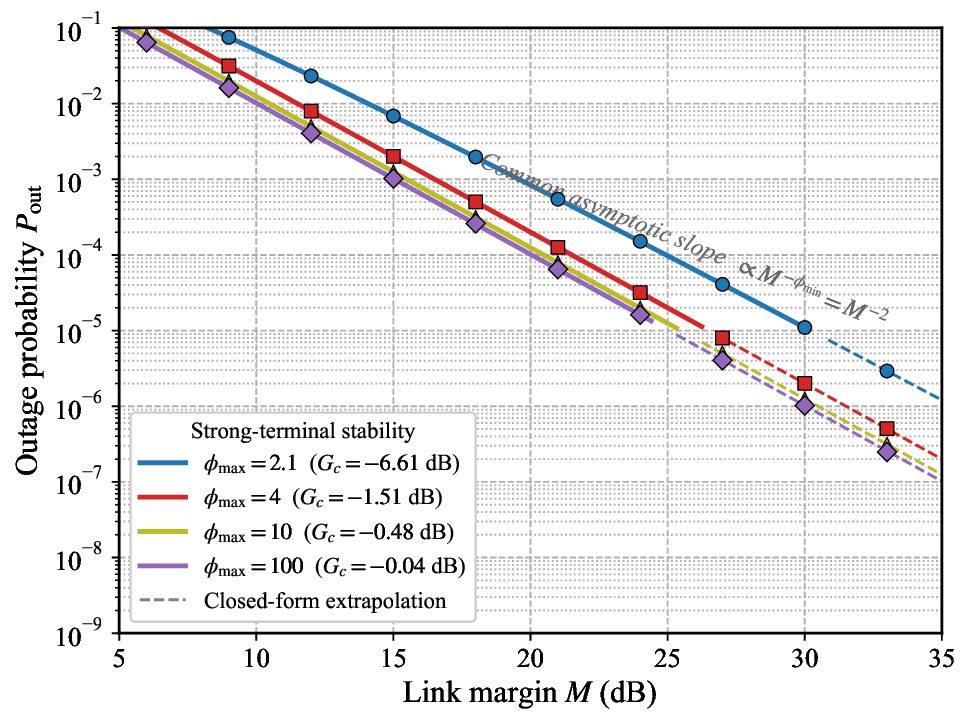}
	\caption{Impact of asymptotic power offset for fixed weakest terminal stability $\phi_{\min}=2$ and varying stronger terminal stability $\phi_{\max}$\rev{, computed from \eqref{outage_simp} with Gaussian--Rayleigh Monte Carlo verification}. All curves exhibit identical asymptotic slopes (the outage decay rate is governed by the weakest link). Increasing $\phi_{\max}$ produces a horizontal shift through the power offset $G_c$, but the associated margin reduction saturates rapidly at high $\phi_{\max}$. }
	\label{fig:outage_slope_gain}
\end{figure}

Fig.~\ref{fig:outage_slope_gain} isolates the effect of the asymptotic power offset for a fixed weakest terminal stability of $\phi_{\min}=2$ by varying the stability of the stronger terminal $\phi_{\max}$. As predicted by the asymptotic analysis, all curves exhibit the same slope on the log scale, confirming that the outage decay rate is entirely governed by the weakest link and remains unchanged as $\phi_{\max}$ increases. Improving the stronger terminal shifts the outage curves leftward, which reflects a reduction in the required link margin through an improved power offset $G_c$. However, this benefit saturates rapidly. Increasing $\phi_{\max}$ from $2.1$ to $4.0$ yields a noticeable horizontal shift, corresponding to several decibels of margin reduction for a given outage target, whereas further increases from $10$ to $100$ produce only marginal additional shifts, with the curves nearly overlapping. This behaviour demonstrates diminishing benefits from strengthening only one terminal and confirms that once the link is constrained by a weak terminal, further improvements at the stronger terminal yield only limited margin savings, while substantial reliability gains require increasing $\phi_{\min}$.

\begin{figure}[htbp]
	\centering
	\includegraphics[width=0.85\columnwidth]{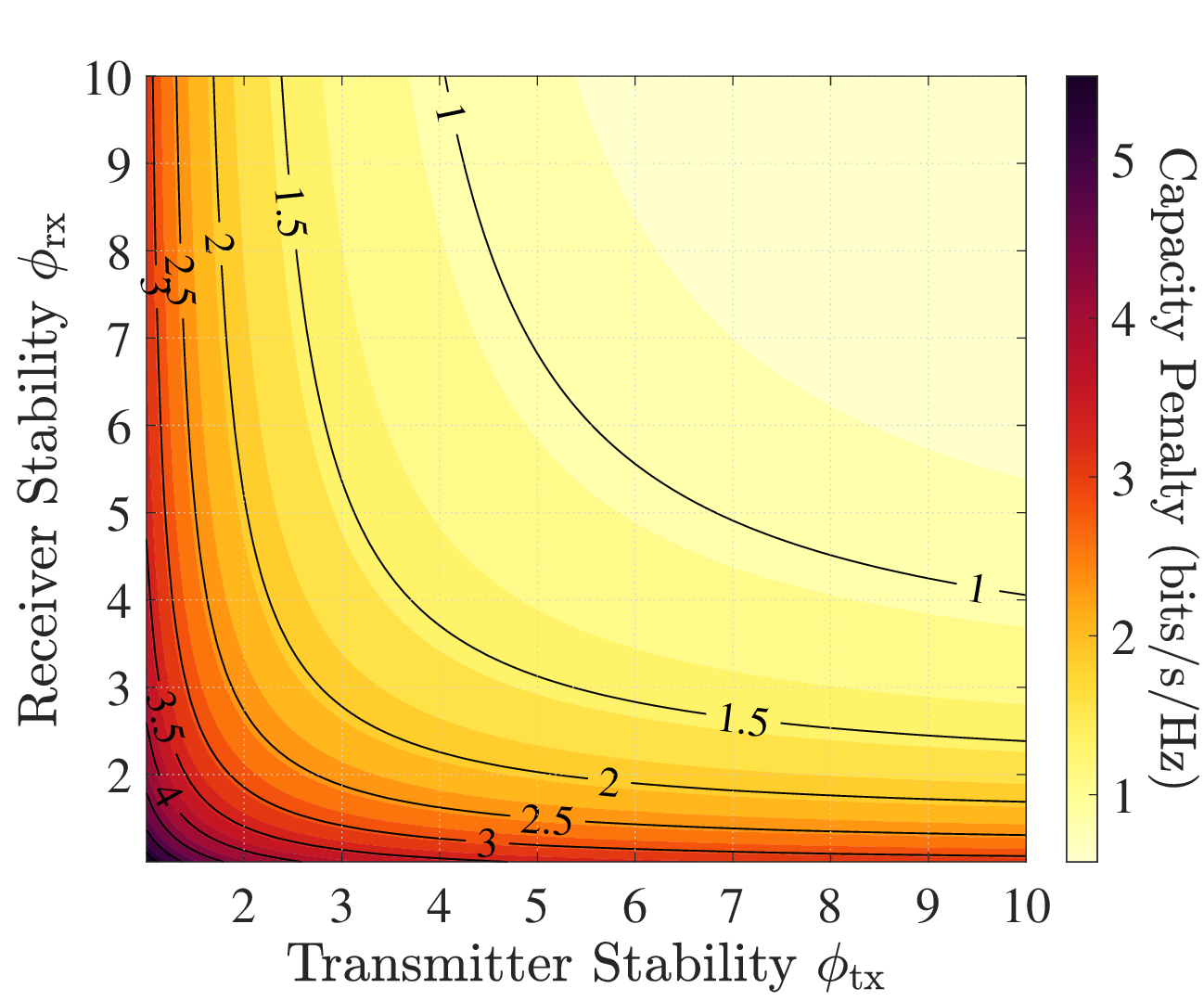}
	\caption{\rev{High-SNR ergodic capacity penalty induced by pointing jitter as a function of the transmitter and receiver stability parameters, $\phi_{\text{tx}}$ and $\phi_{\text{rx}}$, computed from the Gaussian--Rayleigh closed-form expression \eqref{eq:capacity_penalty}.} The contour lines indicate constant capacity loss levels (in bits/s/Hz).}
	\label{fig:capacity_penalty}
\end{figure}

\subsubsection{Ergodic Capacity}
Fig.~\ref{fig:capacity_penalty} illustrates the ergodic capacity penalty induced by pointing jitter as a function of the transmitter and receiver stability parameters, $\phi_{\text{tx}}$ and $\phi_{\text{rx}}$. The contour lines closely follow the analytical relation in \eqref{eq:capacity_penalty}, confirming that the penalty decays with the sum of the inverse stabilities rather than being dominated by the weaker terminal alone. In the low stability regime, where either $\phi_{\text{tx}}$ or $\phi_{\text{rx}}$ approaches unity, the capacity loss increases sharply, exceeding $4$ bits/s/Hz, indicating severe throughput degradation even at high SNR. As both terminals become more stable, the penalty decreases smoothly and becomes very small for $\phi_{\text{tx}}, \phi_{\text{rx}} \gtrsim 6$ \rev{within the Gaussian--Rayleigh model}, where further increases in stability provide only minor improvements in ergodic capacity. The curved contour geometry highlights the coupling between the two terminals, i.e., improvements at one terminal partially compensate for instability at the other, but the strongest reduction in capacity penalty is achieved when both terminals are jointly stabilized. This behaviour complements the outage-based weakest link principle by showing that while link availability is dictated by the weaker terminal, long-term spectral efficiency depends on the combined stability of both ends. \rev{Because the exact Airy-on-detector response decays faster than the curvature-matched Gaussian at moderate displacements (cf.\ Fig.~\ref{fig:gauss_error_db}), the same logarithmic-loss integration applied to the exact response is expected to shift the saturation boundary to somewhat higher $\phi$ values. The qualitative shape of the contour map and the joint-stabilization principle, however, are robust features of the Gaussian--Rayleigh model that carry over to the exact response.}

 Finally, these design principles can be directly operationalized using the \rev{coordinate-descent heuristic} summarized in Algorithm~\ref{design:alg}, which guides the joint selection of beam parameters and link margin under the weakest-link constraint \rev{and is a solver for Problem~\eqref{eq:P1} of Section~\ref{subsec:opt_problem}}.

 \begin{algorithm}
 	\caption{\rev{Coordinate-Descent Heuristic for Problem~\eqref{eq:P1}}}
 	\begin{algorithmic}[1]
 		\label{design:alg}
 		\REQUIRE Link distance $z$, wavelength $\lambda$, target outage $P_{\text{out}}^{\star}$
 		\ENSURE Selected beam divergence $\theta_{\text{div}}$ and required link margin $M$
 		\STATE Obtain pointing jitter estimates $\sigma_{\text{tx}}$ and $\sigma_{\text{rx}}$ from ADCS and PAT specifications
 		\STATE Initialize $\theta_{\text{div}}$ based on diffraction limit and practical beam-shaping constraints
 		\STATE Compute stability parameters $\phi_{\text{tx}}$ and $\phi_{\text{rx}}$ using \eqref{stability_parameters}
 		\STATE Identify the bottleneck terminal $\phi_{\min} = \min(\phi_{\text{tx}}, \phi_{\text{rx}})$
 		\STATE Initialize margin $M$ based on link budget constraints
 		\STATE Compute $P_{\text{out}}$ using \eqref{outage_simp} or \eqref{outage_symetric}
 		\WHILE{$P_{\text{out}} > P_{\text{out}}^{\star}$}
 		\IF{$\phi_{\text{tx}} \gg \phi_{\text{rx}}$}
 		\STATE Increase receiver angular tolerance (e.g., widen $\theta_{\text{fov}}$), or reduce $\sigma_{\text{rx}}$ if feasible (via improved ADCS/tracking)
 		\ELSIF{$\phi_{\text{rx}} \gg \phi_{\text{tx}}$}
 		\STATE Increase transmitter beam divergence $\theta_{\text{div}}$ or reduce $\sigma_{\text{tx}}$ if feasible
 		\ELSE
 		\STATE Adjust $\theta_{\text{div}}$ and $\theta_{\text{fov}}$ jointly to maintain $\phi_{\text{tx}} \approx \phi_{\text{rx}}$
 		\ENDIF
 		\STATE Update $\phi_{\text{tx}}$, $\phi_{\text{rx}}$ and recompute $P_{\text{out}}$
 		\IF{further optical stabilization is infeasible}
 		\STATE Increase margin $M$ and recompute $P_{\text{out}}$
 		\ENDIF
 		\ENDWHILE
 		\RETURN Selected $\theta_{\text{div}}$ and required margin $M$
 	\end{algorithmic}
 \end{algorithm}

\subsection{Beamwidth Optimization Tradeoff}
\label{subsec:beamwidth_opt}
\begin{figure}[htbp]
	\centering
	\includegraphics[width=0.85\columnwidth]{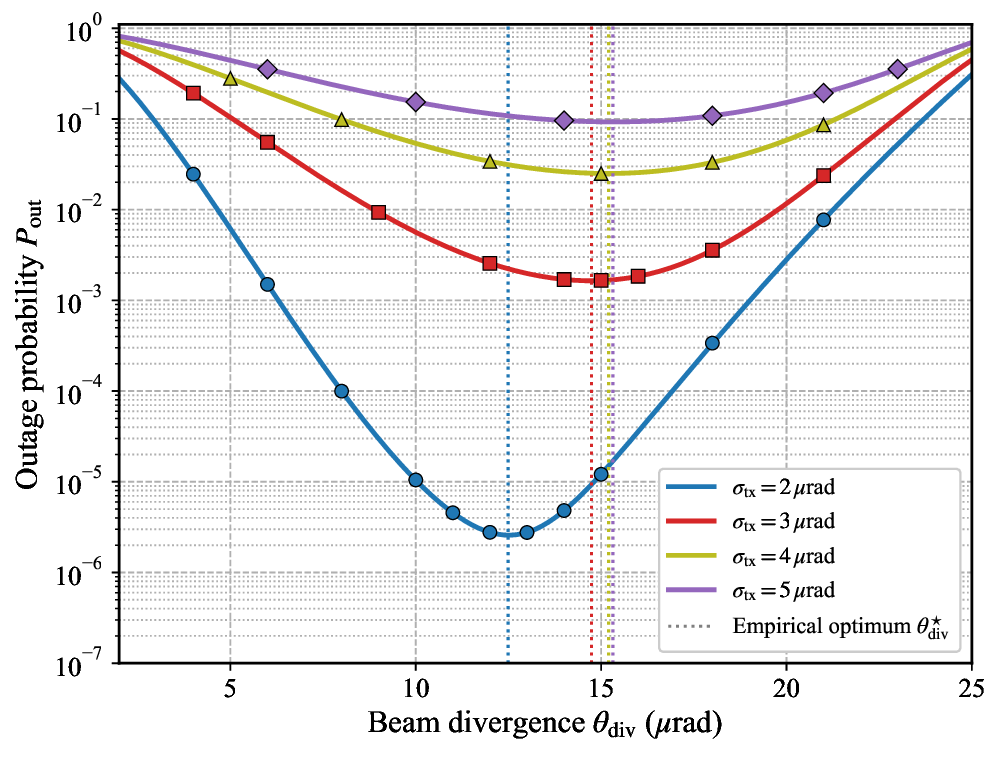}
	\caption{Outage probability versus beam divergence for different transmitter pointing jitter levels\rev{, computed from \eqref{outage_simp} and verified against Monte Carlo of the Gaussian--Rayleigh channel model}. The optimal divergence shifts to larger values as platform jitter increases, highlighting the need to jointly design beamwidth and pointing stability under practical payload and acquisition constraints.  Other simulation parameters: $ D_\textrm{rx}$ = 8 cm, $\theta_{\text{fov}} = 20 \mu \textrm{rad}$, $ \sigma_{\text{rx}} = 3 \mu \textrm{rad}$, $G_\text{th} = 1.2\times10^{-6}$, $\eta_\text{tx} = 0.7, \eta_\text{rx} = 0.7, \eta_{\text{opt}} = 0.5$.}
	\label{fig:beamwidth_optimization}
\end{figure}

Fig.~\ref{fig:beamwidth_optimization} illustrates the tradeoff between beam divergence and pointing jitter in determining OISL outage performance. Narrowing the beam increases the on-axis gain and improves link budget under perfect alignment, but it also amplifies sensitivity to pointing errors, leading to deeper and more frequent fades. Conversely, widening the beam improves robustness to jitter at the cost of reduced peak gain and increased geometric loss. The resulting outage curves therefore exhibit a clear minimum for each jitter level, reflecting the fundamental balance between geometric attenuation and pointing-induced fading.

For each jitter level, there exists an optimal beam divergence that minimizes the outage probability. This optimum shifts monotonically with platform stability. As the transmitter jitter standard deviation increases, the outage-minimizing divergence moves to larger values, for example from approximately $\theta_{\text{div}} \approx 12~\mu\text{rad}$ for $\sigma_{\text{tx}} = 2~\mu\text{rad}$ to $\theta_{\text{div}} \approx 18~\mu\text{rad}$ for $\sigma_{\text{tx}} = 5~\mu\text{rad}$. This trend is captured analytically by the stability parameter $\phi_{\text{tx}}$, which must be selected in conjunction with the beam divergence to achieve the minimum outage\footnote{\rev{The specific optimal divergence values cited above are computed from the Gaussian--Rayleigh model. At the high-jitter operating point $\sigma_\text{tx}=5\,\mu$rad the corresponding stability is $\phi_\text{tx}\approx 3.24$, which falls below the high-stability bound of Section~\ref{subsubsec:approx_accuracy}. Per Table~\ref{tab:exact_mc}, the exact-model optimum may shift by a small fraction of a microradian relative to the value reported here, while the qualitative trend of larger $\sigma_\text{tx}$ requiring wider $\theta_\text{div}$ is preserved.}}.

The performance-optimal divergence must be interpreted in the context of physical payload constraints. In the truncated Gaussian model, the beam divergence and transmitter aperture are linked by $D_{\text{tx}} = (2 f_{\text{trunc}}/\pi)(\lambda/\theta_{\text{div}})$, with $f_{\text{trunc}} \approx 1.48$ for an optimally truncated unobscured aperture \cite{jsac-channel}. Consequently, pushing toward smaller $\theta_{\text{div}}$ requires a larger telescope aperture, directly increasing payload mass, volume, and structural complexity while tightening surface figure and alignment tolerances. In addition, very narrow beams place stringent demands on acquisition and tracking, as the angular search space for beacon detection shrinks, increasing acquisition time and sensitivity to platform uncertainty. As a result, the practically optimal divergence is often chosen slightly larger than the analytically optimal value, trading a small increase in geometric loss for reduced payload burden and more robust acquisition and tracking.

\subsection{Bi-Directional Asymmetric Link}
\label{subsec:Bi-Directional Asymmetric Link}
Consider a LEO--GEO link with $\sigma_{\text{LEO}} = 5$ $\mu$rad, $\sigma_{\text{GEO}} = 1$ $\mu$rad, and identical beamwidths $\theta_{\text{div}} = 10$ $\mu$rad. The forward link (LEO $\to$ GEO) has $\phi_{\text{tx,LEO}} = 1.0$, $\phi_{\text{rx,GEO}} = 25$, \rev{yielding a high-SNR capacity penalty of approximately $3.00$~bits/s/Hz per \eqref{eq:capacity_penalty} (with $\xi=2$), and a high-SNR equivalent-SNR loss of approximately $10\log_{10}e^{-2.08}\approx -9.03$~dB from the corrected \eqref{eq:gamma_eff}}. The return link (GEO $\to$ LEO) has $\phi_{\text{tx,GEO}} = 25$, $\phi_{\text{rx,LEO}} = 1.0$, giving the same penalty. Thus, despite the GEO terminal's high stability, the overall symmetric capacity is still limited by the LEO terminal's reception and transmission. \rev{This LEO--GEO example deliberately operates below the high-stability bound of Section~\ref{subsubsec:approx_accuracy} ($\phi_\text{LEO}=1$); the qualitative conclusion that the symmetric capacity is limited by the weaker terminal is robust to the diffraction approximation, while the absolute capacity-penalty value of $3.00$~bits/s/Hz quoted above is the Gaussian--Rayleigh prediction and should be regarded as an order-of-magnitude indicator. The TX-limited entry of Table~\ref{tab:exact_mc} ($\phi_\text{tx}=1$, $\phi_\text{rx}=20$) provides the closest available calibration point for this regime.}

Optimizing according to \eqref{eq:theta_ratio} would yield $\theta_{\text{div,LEO}} = 5\,\theta_{\text{div,GEO}}$, corresponding to a $50~\mu\text{rad}$ beam for the LEO satellite and a $10~\mu\text{rad}$ beam for the GEO satellite, thereby balancing the stability parameters in both directions. However, widening the LEO beam imposes a transmit-power penalty of $(50/10)^2 = 25$, or approximately 14~dB, to maintain the same link budget. \rev{This is consistent with the corrected power-scaling rule \eqref{eq:Ptx_ratio}, in which the less-stable terminal incurs a $(\sigma_A/\sigma_B)^2$ transmit-power penalty.} This requirement may be prohibitive for a small satellite with limited power resources. In such cases, the designer must either accept a lower stability parameter (and thus a higher outage probability) or invest in improved pointing control to reduce $\sigma_{\text{LEO}}$.

\section{Concluding Remarks and Future Directions}
\label{sec:conclusion}

In this paper, we studied the statistical performance of optical inter-satellite links under platform-induced pointing jitter and developed accurate and computationally efficient closed-form models for the end-to-end channel gain, outage probability, and ergodic capacity penalty. By applying Gaussian main-lobe approximations to both the transmitter far-field intensity pattern and the receiver coupling efficiency, the diffraction-based channel response was transformed into tractable expressions that enable direct link-margin evaluation across both stable and jitter-limited operating regimes. The results demonstrate that OISL reliability is fundamentally governed by terminal stability rather than transmit power alone. In particular, the asymptotic decay of the outage probability is dictated by the weaker terminal, with $d = \phi_{\min} = \min(\phi_{\text{tx}}, \phi_{\text{rx}})$, while improving only the stronger terminal affects the power offset without changing the fundamental slope of the reliability curve. Consequently, balancing the pointing stability of the two terminals is substantially more effective than over-engineering a single end, as terminal imbalance introduces a fixed margin penalty with rapidly diminishing returns. The beamwidth optimization results further highlight that, although narrower beams increase geometric gain, they exacerbate sensitivity to pointing errors and require larger transmit apertures, leading to nontrivial payload and acquisition tradeoffs. Hence, the proposed analytical framework and design methodology enable a direct translation of ADCS and PAT capabilities into availability targets, providing a principled basis for robust dimensioning of future high-capacity OISL systems.

Several directions for future work are of interest. \rev{The closed-form expressions developed here cover the canonical zero-mean Rayleigh-jitter regime and the high-margin asymptotic limit. Quantitative sensitivity sweeps over the anisotropy parameter $q$ (Hoyt) and the Rician $K$-factor of Section~\ref{sec:rayleigh_extensions} are left as future work, as is a specific correlated-jitter model (common-mode/differential-mode decomposition) that would tighten the Fr\'echet--Hoeffding bound envelope \eqref{eq:frechet_hoeffding} for bidirectional links. Additional directions include}: \rev{extending the present framework to incorporate} temporally correlated pointing jitter and closed-loop tracking dynamics, as well as non-Gaussian beam profiles and partially coherent optical sources\rev{; modeling} acquisition and re-acquisition phases under large initial misalignment, where the Gaussian approximation may no longer be accurate\rev{, requiring direct use of the exact Klein--Degnan/Airy responses}; and \rev{extending} the analysis to multi-hop inter-satellite networks and constellation-level performance metrics, such as end-to-end availability and routing-aware outage, \rev{which} would provide further insight into the design of large-scale optical satellite networks.

\section*{Acknowledgement}
{The authors acknowledge support by the European Space Agency (ESA) under grants no. 1000038755-8000022976-1 ``Development of a High Speed Optical	Inter Satellite Link (OISL) Terminal
	Based on a Low Complexity Platform.''}

\balance

\bibliographystyle{IEEEtraN}

\end{document}